\documentclass[11pt,twoside]{article}
\usepackage{amsmath}
\usepackage{amssymb}
\def\EndBox#1{
\hskip0.1em\hfill\null\ \null\nobreak\hfill\kern3pt
\hbox{$\scriptstyle #1$} \smallbreak}
\def\qed{\EndBox{\square}}
\newtheorem{lemma}{Lemma}[section]
\newtheorem{proposition}{Proposition}[section]
\newcommand{\proof}{{\sc proof:~}}
\newcommand{\remark}{\smallbreak\noindent{\bf Remark.}}
\newcommand{\dbk}{\displaybreak[2]\\}
\renewcommand{\a}{\alpha}
\renewcommand{\b}{\beta}
\newcommand{\g}{\gamma}
\newcommand{\G}{\Gamma}
\renewcommand{\d}{\delta}
\newcommand{\e}{\varepsilon}
\newcommand{\z}{{\zeta}}

\renewcommand{\th}{\theta}
\newcommand{\Th}{\Theta}
\newcommand{\vth}{\vartheta}
\renewcommand{\l}{\lambda}
\newcommand{\m}{\mu}
\newcommand{\n}{\nu}
\renewcommand{\r}{\rho}
\newcommand{\s}{\sigma}
\renewcommand{\t}{\tau}

\newcommand{\om}{{\omega}}
\newcommand{\Cs}{
	{\rlap{\lower3pt\hbox{\textnormal{\LARGE\char'040}}}{\Gamma}}{}}
\newcommand{\de}{\partial}
\newcommand{\oh}{\tfrac{1}{2}}
\newcommand{\ih}{\tfrac{\iO}{2}}
\newcommand{\oq}{\tfrac{1}{4}}
\newcommand{\iq}{\tfrac{\iO}{4}}
\newcommand{\osq}{\tfrac{1}{\surd2}}
\newcommand{\isq}{\tfrac{\iO}{\surd2}}

\newcommand{\cj}[1]{\overline{#1}}
\newcommand{\lin}{{\scriptscriptstyle\bigstar}}
\newcommand{\alin}{{\overline{\scriptscriptstyle\bigstar}}}
\renewcommand{\.}{{\scriptstyle\boldsymbol{\dot{}}}}
\newcommand{\bul}{\bullet}
\newcommand{\td}{\tilde}
\newcommand{\tb}[1]{\Tilde{\Bar{#1}}}
\newcommand{\up}{{\scriptscriptstyle\uparrow}}
\newcommand{\grav}{{}_{\mathrm{g}}}
\newcommand{\emag}{{}_{\mathrm{em}}}
\newcommand{\Dir}{{}_{\sst{\mathrm{D}}}}
\newcommand{\ost}[1]{\overset{{}_{{\,}_*}}#1}
\newcommand{\fl}{\flat}
\newcommand{\bp}{\backprime}
\newcommand{\bl}{{\bar\lambda}}
\newcommand{\bh}{{\bar h}}
\newcommand{\bq}{{\bar q}}
\newcommand{\bs}{{\bar s}}
\newcommand{\bu}{{\bar u}}
\newcommand{\bv}{{\bar v}}
\newcommand{\bw}{{\bar w}}
\newcommand{\bzz}{{\bar\zz}}
\newcommand{\be}{{\bar\varepsilon}}
\newcommand{\bze}{{\bar\zeta}}
\newcommand{\bch}{{\bar\chi}}
\newcommand{\A}{{\boldsymbol{A}}}
\newcommand{\C}{{\boldsymbol{C}}}
\newcommand{\D}{{\boldsymbol{D}}}
\renewcommand{\H}{{\boldsymbol{H}}}
\newcommand{\J}{{\boldsymbol{J}}}
\newcommand{\K}{{\boldsymbol{K}}}
\newcommand{\M}{{\boldsymbol{M}}}
\newcommand{\N}{{\boldsymbol{N}}}
\newcommand{\Q}{{\boldsymbol{Q}}}
\newcommand{\Ql}{\Q{}^\lin}
\newcommand{\Qa}{\cj{\Q}{}^\lin}
\let\Sec=\S
\renewcommand{\S}{{\boldsymbol{S}}}
\newcommand{\Sc}{\cj{\S}}
\newcommand{\Sl}{\S{}^\lin}
\newcommand{\U}{{\boldsymbol{U}}}
\newcommand{\Uc}{\cj{\U}}
\newcommand{\Ua}{\cj{\U}{}^\lin}
\newcommand{\Ul}{\U{}^\lin}
\newcommand{\V}{{\boldsymbol{V}}}
\newcommand{\Vc}{\cj{\V}}
\newcommand{\Val}{\V{}^\alin}
\newcommand{\Va}{\cj{\V}{}^\lin}
\newcommand{\Vl}{\V{}^\lin}
\newcommand{\W}{{\boldsymbol{W}}}
\newcommand{\Wc}{\cj{\W}}

\newcommand{\Wl}{\W{}^\lin}
\newcommand{\Lie}{\mathfrak{L}}
\newcommand{\Ug}{\mathrm{U}}
\newcommand{\Gl}{\mathrm{Gl}}
\newcommand{\SlG}{\mathrm{Sl}}
\newcommand{\SU}{\mathrm{SU}}
\newcommand{\Lor}{\mathrm{Lor}}
\newcommand{\Cl}{\mathrm{Cl}}
\newcommand{\Pin}{\mathrm{Pin}}
\newcommand{\Spin}{\mathrm{Spin}}
\newcommand{\CC}{{\mathbb{C}}}
\newcommand{\LL}{{\mathbb{L}}}
\newcommand{\QQ}{{\mathbb{Q}}}
\newcommand{\RR}{{\mathbb{R}}}
\newcommand{\UU}{{\mathbb{U}}}
\newcommand{\ZZ}{{\mathbb{Z}}}
\newcommand{\CCc}{{\scriptscriptstyle{\mathbb{C}}}}
\newcommand{\RRr}{{\scriptscriptstyle{\mathbb{R}}}}
\newcommand{\Ccal}{{\mathcal{C}}}
\newcommand{\Ecal}{{\mathcal{E}}}
\newcommand{\Lcal}{{\mathcal{L}}}
\newcommand{\Pcal}{{\mathcal{P}}}
\newcommand{\Tcal}{{\mathcal{T}}}
\newcommand{\End}{\operatorname{End}}
\newcommand{\Aut}{\operatorname{Aut}}
\newcommand{\Tr}{\operatorname{Tr}}
\newcommand{\Ad}{\operatorname{Ad}}
\newcommand{\Id}[1]{{1\!\!1}\!{}_{#1}{}}
\newcommand{\id}{{1\!\!1}}
\newcommand{\dO}{\mathrm{d}}
\newcommand{\KO}{\mathrm{K}}
\newcommand{\kO}{\mathrm{k}}
\newcommand{\pO}{\mathrm{p}}
\newcommand{\tdpO}{\tilde{\mathrm{p}}}
\newcommand{\TO}{\mathrm{T}}
\newcommand{\TS}{\TO^{*}\!}
\newcommand{\VO}{\mathrm{V}}
\newcommand{\dx}{\dO\xx}
\newcommand{\eO}{\mathrm{e}}
\newcommand{\iO}{\mathrm{i}}
\newcommand{\na}{\nabla\!}
\newcommand{\nasl}{{\rlap{\raise1pt\hbox{\,/}}\nabla}}
\newcommand{\ten}[1]{\operatorname*{\otimes}_{\!{\scriptscriptstyle #1}} }
\newcommand{\dir}[1]{\operatorname*{\oplus}_{\!{\scriptscriptstyle #1}} }
\newcommand{\we}{{\,\wedge\,}}
\newcommand{\weu}[1]{{\wedge^{\!#1}}}
\newcommand{\pint}{\mathord{\rfloor}}
\newcommand{\comp}{\mathbin{\raisebox{1pt}{$\scriptstyle\circ$}}}
\newcommand{\tn}{{\,\otimes\,}}
\newcommand{\vh}{{\,\bar\vee\,}}
\newcommand{\bang}[1]{{\langle#1\rangle}}
\newcommand{\pmap}{{(\!{+}\!)}}
\newcommand{\mmap}{{(\!{-}\!)}}
\newcommand{\Ii}[2]{{}^{#1}_{\phantom{#1}\!#2}}
\newcommand{\iI}[2]{{}_{#1}^{\phantom{#1}\!#2}}
\newcommand{\iIi}[3]{{}_{#1\phantom{#2}\!\!#3}^{\phantom{#1}\!#2}}
\newcommand{\si}[2]{\sigma\iI{#1}{#2}}
\newcommand{\ti}[2]{\tau\iI{#1}{#2}}
\newcommand{\sA}{{\scriptscriptstyle A}}
\newcommand{\sB}{{\scriptscriptstyle B}}
\newcommand{\sC}{{\scriptscriptstyle C}}
\newcommand{\sD}{{\scriptscriptstyle D}}
\newcommand{\cA}{{\sA\.}}
\newcommand{\cB}{{\sB\.}}
\newcommand{\cC}{{\sC\.}}
\newcommand{\cD}{{\sD\.}}
\newcommand{\AAd}{{\sA\cA}}
\newcommand{\BBd}{{\sB\cB}}
\newcommand{\zeA}{{\zeta_\sA}}
\newcommand{\bzeA}{{\bze_\cA}}
\newcommand{\zeB}{{\zeta_\sB}}
\newcommand{\bzeB}{{\bze_\cB}}
\newcommand{\zzA}{\zz^\sA}
\newcommand{\zzB}{\zz^\sB}
\newcommand{\bzzA}{\bzz^\cA}
\newcommand{\bzzB}{\bzz^\cB}
\newcommand{\bb}{{\mathsf{b}}}
\newcommand{\Ksf}{{\mathsf{K}}}
\newcommand{\ee}{{\mathsf{e}}}
\renewcommand{\tt}{{\mathsf{t}}}
\newcommand{\ww}{{\mathsf{w}}}

\newcommand{\xx}{{\mathsf{x}}}
\newcommand{\zz}{{\mathsf{z}}}
\newcommand{\ie}{i.e$.$}

\newcommand{\phexp}{{\phantom{a}}}
\newcommand{\hm}{\phantom{-}}
\newcommand{\sst}{\scriptscriptstyle}
\newcommand{\sTh}{{\breve\Theta}}
\newcommand{\qRq}{{\quad\Rightarrow\quad}}
\newcommand{\cliffordplusmatrix}[1]{
	\left(\begin{smallmatrix}#1&0\\ 0&#1^\ddag\end{smallmatrix}\right)}
\newcommand{\sKM}{
	\left(\begin{smallmatrix}K&0\\ 0&K^\ddag\end{smallmatrix}\right)}
\newcommand{\sPM}{
	\left(\begin{smallmatrix}0&P\\ P^\ddag&0\end{smallmatrix}\right)}
\title{``Minimal geometric data'' approach to
Dirac algebra, spinor groups and field theories}
\date{{\small 2nd revised version, 13 February 2008} }
\author{Daniel Canarutto\\[6pt]
{\small\it Dipartimento di Matematica Applicata ``G. Sansone'', }\\
{\small\it Via S. Marta 3, 50139 Firenze, Italia}\\
{\small http://www.dma.unifi.it/\char126 canarutto}}
\begin{document}
\bibliographystyle{unsrt}
\maketitle
\begin{abstract}\noindent
The three first sections contain an updated, not-so-short account
of a partly original approach to spinor geometry and field theories
introduced by Jadczyk and myself~\cite{CJ97b,C98,C00b};
it is based on an intrisic treatment of 2-spinor geometry
in which the needed background structures have not to be assumed,
but rather arise naturally from a unique geometric datum:
a vector bundle with complex 2-dimensional fibres
over a real 4-dimensional manifold.
The two following sections deal with Dirac algebra and 4-spinor groups
in terms of two spinors,
showing various aspects of spinor geometry from a different perspective.
The last section examines particle momenta in 2-spinor terms
and the bundle structure of 4-spinor space over momentum space.
\end{abstract}

\noindent
AMS 2000 MSC:
15A66, 
83C22. 

\smallbreak\noindent
Keywords:\\
Dirac algebra,
spinor groups,
two-spinors,
Einstein-Cartan-Maxwell-Dirac fields

\smallbreak\noindent
{\bf Acknowledgements:} Thanks are due to Carlo Franchetti
(Dipartimento di Ma\-te\-ma\-ti\-ca Applicata ``G. Sansone'', Firenze)
for useful discussions and suggestions.
\tableofcontents
\section*{Introduction}

The precise equivalence between the $4$-spinor and $2$-spinor settings
for electrodynamics was exposed by Jadczyk and myself
in~\cite{CJ97a,CJ97b,C98,C00b}.
In summary one sees that, from an algebraic point of view,
the only notion of a complex $2$-dimensional vector space $\S$ yields,
naturally and without any further assumptions,
all the needed algebraic structures through functorial constructions;
conversely in a $4$-spinor setting, provided one makes the minimum assumptions
which are needed in order to formulate the standard physical theory,
the $4$-spinor space naturally splits (Weyl decomposition)
into the direct sum of two 2-dimensional subspaces
which are anti-dual to each other.
In a sense, which setting one regards as fundamental is then
mainly a matter of taste.
The $4$-spinor setting is closer to standard notations,
and some formulas can be written in a more compact way,
while the relations among the various objects are somewhat more involved.
The $2$-spinor setting
turns out to give a much more direct formulation,
in which all the basic objects and the relations among them
naturally set into their places;
just from $\S$ one authomatically gets
\emph{exactly} the needed algebraic structure, nothing more, nothing less:
4-spinor space $\W$ with the `Dirac adjoint' anti-isomorphism,
Minkowski space $\H$ and Dirac map $\g:\H\to\End(\W)$
with the required properties.
Further objects which are commonly considered
depend on the choice of a gauge of some sort,
whose nature is precisely described.

When we consider a vector bundle $\S\to\M$,
where now the fibres are complex 2-dimensional
and $\M$ is a real 4-dimensional manifold,
then we don't have to assign any further background structure
in order to formulate a full Einstein-Cartan-Maxwell-Dirac theory.
In fact we naturally get a vector bundle $\H\to\M$
whose fibres are Minkowski spaces, a 4-spinor bundle $\W\to\M$ and so on.
Any object which is not determined by geometric construction
from the unique geometric datum $\S\to\M$ is a \emph{field} of the theory,
namely we consider: the tetrad $\Th:\TO\M\to\LL\tn\H$,
the 2-spinor connection $\Cs$, the electromagnetic and Dirac fields.
(Even coupling factors naturally arise as covariantly constant sections
of the real line bundle $\LL$ of \emph{length units},
which is geometrically constructed from $\S$.)
The gravitational field is described by the tetrad
(which can be seen as a `square root' of spacetime metric)
and by the connection induced by $\Cs$ on $\H$,
while the remaining part of the spinor connection
can be viewed as the electromagnetic potential.
A natural Lagrangian density for all these fields is then introduced;
the relations between metric and connection
and between e.m.\ potential and e.m.\ field follow from the
(Euler-Lagrange) field equations.
All considered, this setting has some original aspects
but is not in contrast to the (by now classical) Penrose formalism~\cite{PR84}.

In~\Sec\ref{S:Dirac algebra in two-spinor terms}
and~\Sec\ref{S:Clifford group and its subgroups}
I'll show how the above said algebraic setting,
and in particular the natural splitting of the 4-spinor space
into the direct sum of its Weyl subspaces,
enables us to examine the structures of the Dirac algebra, the Clifford group
and its subgroups from a different perspective.

In~\Sec\ref{S:Spinors and particle momenta}
I'll show the strict relation existing between the two-spinor setting
and the geometry of particle momenta,
in particular the bundle structure of $\W$ over the space of momenta.
These results 
are a preparation to a 2-spinor formulation
of quantum electrodynamics along le lines of a previous paper~\cite{C05},
in which the classical structure underlying electron states
is a 2-fibred bundle over spacetime.

\section{Two-spinor geometry}
\label{S:Two-spinor geometry}
In this section we'll see how all the fundamental geometric structures
needed for Dirac theory naturally arise through functorial constructions
from a two-dimensional complex vector space,
with no further assumptions.

\subsection{Complex conjugated spaces}\label{s:Complex conjugated spaces}

If $\A$ is a set and $f:\A\to\CC$ is any map,
then $\bar f:\A\to\CC:a\mapsto\overline{f(a)}$ is the conjugated map.
Let $\V$ be a complex vector space of finite-dimension $n$\,;
its \emph{dual} space $\Vl$ and \emph{antidual} space $\V^\alin$
are defined to be the $n$-dimensional complex vector spaces
of all maps $\V\to\CC$ which are respectively linear and antilinear.
One then has the distinguished anti-isomorphism $\Vl\to\V^\alin:\l\mapsto\bl$\,.

Set now $\Vc:=\V^{\lin\alin}$, and call this the \emph{conjugate space} of $\V$.
One has the natural isomorphisms
$$\V\cong\V^{\lin\lin}\cong\V^{\alin\,\alin}~,\quad
\Vc:=\V^{\lin\alin}\cong\V^{\alin\lin}~.$$
Summarizing, one one gets the four distinct spaces
$$\V\leftrightarrow\Vc~,\quad \Vl\leftrightarrow\Val~,$$
where the arrows indicate the conjugation anti-isomorphisms.

Accordingly, coordinate expressions have four types of indices.
Let $(\bb_\sA)$, $1\leq A\leq n$\,, be a basis of $\V$
and $(\bb^{\sA})$ its dual basis of $\Vl$.
The corresponding indices in the conjugate spaces are distinguished by a dot,
namely one writes
$$\bar\bb_{\cA}:=\overline{\bb_\sA}~,\quad \bar\bb^{\cA}
:=\overline{\bb^{\sA}}~,$$
so that $\{\bar\bb_{\cA}\}$ is a basis of $\Vc$ and
$\{\bar\bb^{\cA}\}$ its dual basis of $\Va$.
For $v\in\V$ and $\l\in\Vl$ one has
\begin{align*}&
v=v^\sA\,\bb_\sA~,\quad \bv=\bv^\cA\,\bar\bb_\cA~,\\&
\l=\l_\sA\,\bb^\sA~,\quad \bl=\bl_\cA\,\bar\bb^{\cA}~,
\end{align*}
where $\bv^\cA=\overline{v^\sA}$, $\bl_\cA:=\overline{\l_\sA}$
and Einstein summation convention is used.

The conjugation morphism can be extended to tensors of any rank and type;
if $\t$ is a tensor then all indices of $\bar\t$
are of reversed (dotted/non-dotted) type;
observe that dotted indices cannot be contracted with non-dotted indices.
In particular if $K\in\Aut(\V)\subset\V\tn\Vl$
then $\bar K\in\Aut(\Vc)\subset\Vc\tn\Va$ is the induced
conjugated transformation
(under a basis transformation,
dotted indices transform with the conjugate matrix).

\subsection{Hermitian tensors}\label{s:Hermitian tensors}

The space $\V\tn\Vc$ has a natural real linear (complex anti-linear) involution
$w\mapsto w^\dag$,
which on decomposable tensors reads
$$(u\tn\bv)^\dag=v\tn\bu~.$$
Hence one has the natural decomposition of $\V\tn\Vc$ into the direct sum
of the \emph{real} eigenspaces of the involution with eigenvalues $\pm1$,
respectively called the \emph{Hermitian} and \emph{anti-Hermitian} subspaces,
namely
$$\V\tn\Vc=(\V\vh\Vc)\oplus \iO\,(\V\vh\Vc)~.$$
In other terms, the Hermitian subspace $\V\vh\Vc$ is constituted by
all $w\in\V\tn\Vc$ such that $w^\dag=w$,
while an arbitrary $w$ is uniquely decomposed into the sum of an Hermitian
and an anti-Hermitian tensor as
$$w=\oh(w+w^\dag)+\oh(w-w^\dag)~.$$
In terms of components in any basis,
$w=w^{\sA\cB}\bb_\sA\tn\bar\bb_\cB$ is Hermitian (anti-Hermitian)
iff the matrix $(w^{\sA\cB}\,)$ of its components is of the same type,
namely $\bw^{\cB\sA}=\pm w^{\sA\cB}$.

Obviously $\Vl\tn\Va$ decomposes in the same way,
and one has the natural isomorphisms
$$(\V\vh\Vc)^*\cong\Vl\vh\Va~~,\quad (\iO\,\V\vh\Vc)^*\cong\iO\,\Vl\vh\Va~,$$
where ${}^*$ denotes the \emph{real} dual.

A Hermitian $2$-form is defined to be a Hermitian tensor $h\in\Va\vh\Vl$.
The associated quadratic form $v\mapsto h(v,v)$ is real-valued.
The notions of signature and non-degeneracy of Hermitian $2$-forms
are introduced similarly to the case of real bilinear forms.
If $h$ is non-degenerate then it yields the isomorphism
$h^\fl:\Vc\to\Vl:\bv\mapsto h(\bv,\_)$;
its conjugate map is an anti-isomorphism $\Vc\to\Va$ which,
via composition with the canonical conjugation, can be seen as the isomorphism
$\bh^\fl:\V\to\Va:v\mapsto h(\_,v)$.
The inverse isomorphisms $h^\#$ and $\bh^\#$ are similarly related
to a Hermitian tensor $h^{-1}\in\Vc\vh\V$.
One has the coordinate expressions
\begin{align*}
&(h^\fl(\bv))_\sB=h_{\cA\sB}\bv^\cA~~,\quad
&(\bh^\fl(v))_\cA=h_{\cA\sB}v^\sB=\bh_{\sB\cA}\,v^\sB~~,\\
&(h^\#(\bl))^\sB=h^{\cA\sB}\bl_\cA~~,\quad
&(\bh^\#(\l))^\cA=h^{\cA\sB}\l_\sB=\bh^{\sB\cA}\l_\sB~~,
\end{align*}
where $h^{\cC\sA}h_{\cC\sB}=\d\Ii{\sA}{\sB}$\,,
$h^{\cA\sC}h_{\cB\sC}=\d\Ii{\cA}{\cB}$\,.

\subsection{Two-spinor space}\label{s:Two spinor space}

Let $\S$ be a $2$-dimensional complex vector space.
Then $\weu{2}\S$ is a $1$-dimensional complex vector space;
its dual space $(\weu{2}\S)^\lin$ will be identified with
$\weu{2}\Sl$ via the rule\footnote{
Here, $s{\we}s':=\oh(s{\otimes}s'{-}s'{\otimes}s)$\,.
This contraction, defined in such a way to respect
usual conventions in two-spinor literature,
corresponds to $1/4$ standard exterior-algebra contraction.}
$$\om(s{\we}s'):=\oh\om(s,s')~, \quad\forall~\om\in\weu{2}\Sl,~s,s'\in\S~.$$
Any $\om\in\weu2\Sl\setminus\{0\}$
(a `symplectic' form on $\S$)
has a unique `inverse' or `dual' element $\om^{-1}$\,.
Denoting by $\om^\fl:\S\to\Sl$ the linear map defined by
$\bang{\om^\fl(s),t}:=\om(s,t)$
and by $\om^\#:\Sl\to\S$ the linear map defined by
$\bang{\m,\om^\#(\l)}:=\om^{-1}(\l,\m)$\,,
one has $$\om^\#=-(\om^\fl)^{-1}~.$$

The Hermitian subspace of $(\weu{2}\S)\tn(\weu{2}\Sc)$
is a 1-dimensional real vector space with a distinguished orientation,
whose positively oriented semispace
$$\LL^2:=[(\weu{2}\S)\vh(\weu{2}\Sc)]^{+}:=\{w\tn\bw,~w\in\weu{2}\S\}$$
has the square root semi-space $\LL$,
called the space of \emph{length units}.\footnote{
A \emph{unit space} is defined to be a 1-dimensional real semi-space,
namely a positive semi-field associated with the semi-ring $\RR^+$
(see~\cite{CJM,CJ97a} for details).
The \emph{square root} $\UU^{1/2}$ of a unit space $\UU$,
is defined by the condition that $\UU^{1/2}\tn\UU^{1/2}$ be isomorphic to $\UU$.
More generally, any \emph{rational power} of a unit space
is defined up to isomorphism
(negative powers correspond to dual spaces).
In this article we only use the unit space $\LL$ of lengths and its powers;
essentially, this means that we take $\hbar=c=1$\,.} 

Next, consider the complex $2$-dimensional space
$$\U:=\LL^{-1/2}\tn\S~.$$
This is our \emph{$2$-spinor space}.
Observe that the $1$-dimensional space
$$\Q:=\weu{2}\U=\LL^{-1}\tn\weu{2}\S$$
has a distinguished Hermitian metric,
defined as the unity element in
$$\Qa\vh\Ql\equiv(\weu{2}\Ua)\vh(\weu{2}\Ul)
=\LL^{-2}\tn(\weu{2}\Sl)\vh(\weu{2}\Sl)\cong\RR~.$$
Hence there is the distinguished set of normalized symplectic forms on $\U$,
any two of them differing by a phase factor.\footnote{
One says that
elements of $\U$ and of its tensor algebra are `conformally invariant',
while tensorializing by $\LL^r$ one obtains `conformal densities'
of weight $r$.} 

Consider an arbitrary basis
$(\xi_\sA)$ of $\S$ and its dual basis $(\xx^\sA)$ of $\Sl$.
This determines the mutually dual bases
$$\ww:=\e^{\sA\sB}\,\xi_\sA\we\xi_\sB~,\quad
\ww^{-1}:=\e_{\sA\sB}\,\xx^\sA\we\xx^\sB~,$$
respectively of $\weu{2}\S$ and $\weu{2}\Sl$
(here $\e^{\sA\sB}$ and $\e_{\sA\sB}$ both denote
the antisymmetric Ricci matrix),
and the basis 
$$l:=\sqrt{\ww\tn\bar\ww} \quad\text{of}\quad \LL~.$$
Then one also has the induced mutually dual, {\em normalized\/} bases
$$(\zeA):=(l^{-1/2}\tn\xi_\sA)~,\quad (\zzA):=(l^{1/2}\tn\xx^\sA)$$
of $\U$ and $\Ul$, and also
\begin{align*}
&\e:=l\tn\ww^{-1}=\e_{\sA\sB}\,\zzA\we\zzB\in\Ql\equiv\weu{2}\Ul~, \\[6pt]
&\e^{-1}\equiv l^{-1}\tn\ww=\e^{\sA\sB}\,\zeA\we\zeB\in\Q\equiv\weu{2}\U~.
\end{align*}
\remark~
In contrast to the usual $2$-spinor formalism,
no symplectic form is fixed.
The  $2$-form $\e$ is unique up to a phase factor
which depends on the chosen 2-spinor basis,
and determines isomorphisms
\begin{align*}
& \e^\fl:\U\to\Ul:u\mapsto u^\fl~,~~
\bang{u^\fl,v}:=\e(u,v)\qRq (u^\fl)_\sB=\e_{\sA\sB}\,v^\sA~,\\[6pt]
& \e^\#:\Ul\to\U:\l\mapsto\l^\#~,~~
\bang{\m,\l^\#}:=\e^{-1}(\l,\m)\qRq (\l^\#)^\sB=\e^{\sA\sB}\,\l_\sA~.
\end{align*}
If no confusion arises, we'll make the identification $\e^\#\equiv\e^{-1}$.
\smallbreak

\subsection{2-spinors and Minkowski space}
\label{s:2-spinors and Minkowski space}

Though a normalized element $\e\in\Ql$ is unique only up to a phase factor,
certain objects which can be expressed through it
are natural geometric objects.
The first example is the unity element in $\Ql\tn\Qa$,
which can be written as $\e\tn\be$\,;
it can also be seen as a bilinear form $g$ on $\U\tn\Uc$,
given for decomposable elements by
$$g(p\tn\bq,r\tn\bs)=\e(p,r)\,\be(\bq,\bs)~.$$
The fact that any $\e$ is non-degenerate implies that $g$
is non-degenerate too.
In a normalized 2-spinor basis $(\zeA)$ one writes
$w=w^{\AAd}\,\zeA\tn\bzeA\in\U\tn\Uc$,
$g_{\AAd\,\BBd}=\e_{\sA\sB}\,\be_{\cA\cB}$ and\footnote{
Note how $\det w\equiv\det\bigl(w^{\AAd}\:\bigr)$ is intrinsically defined
through $\e$\,, even if $w$ is not an endomorphism.} 
$$g(w,w)=\e_{\sA\sB}\,\be_{\cA\cB}\,w^{\AAd}\,w^{\BBd}=2\,\det w~.$$

Next, consider the Hermitian subspace
$$\H:=\U\vh\Uc\subset\U\tn\Uc~.$$
This is a $4$-dimensional \emph{real} vector space;
for any given normalized basis $(\zeA)$ of $\U$ consider, in particular,
the \emph{Pauli basis} $(\t_\l)$ of $\H$ associated with $(\zeA)$,
namely
$$\t_\l\equiv\t\iI{\l}{\AAd}\,\zeA\tn\bzeA
\equiv\osq\,\si{\l}{\AAd}\,\zeA\tn\bzeA~,\quad \l=0,1,2,3~,$$
where $(\si{\l}{\AAd})$ denotes the $\l$-th Pauli matrix.\footnote{
$\quad\s_0:=\left(\begin{smallmatrix}~1&~\hm0~\\[5pt]
~0&~\hm1~\end{smallmatrix}\right)~,\quad
\s_1:=\left(\begin{smallmatrix}~0&~\hm1~\\[5pt]
~1&~\hm0~\end{smallmatrix}\right)~,\quad
\s_2:=\left(\begin{smallmatrix}~0&~-\iO~\\[5pt]
~\iO&~\hm0~\end{smallmatrix}\right)~,\quad
\s_3:=\left(\begin{smallmatrix}~1&~\hm0~\\[5pt]
~0&~-1~\end{smallmatrix}\right)~.$ } 

The restriction of $g$ to the Hermitian subspace $\H$
turns out to be a Lorentz metric with signature $(+,-,-,-)$\,.
Actually, a Pauli basis is readily seen to be orthonormal, namely
$g_{\l\m}:=g(\t_\l\,,\t_\m)=\eta_{\l\m}:=2\,\d^0_\l\d^0_\m-\d_{\l\m}$\,.

It's not difficult to prove:
\begin{proposition}
An element $w\in\U\tn\Uc=\CC\tn\H$ is null, that is $g(w,w)=0$\,,
iff it is a decomposable tensor: $w=u\tn\bs$, $u,s\in\U$\,.
\end{proposition}

A null element in $\U\tn\Uc$ is also in $\H$
iff it is of the form $\pm u\tn\bu$.
Hence the \emph{null cone} $\N\subset\H$
is constituted exactly by such elements.
Note how this fact yields a way of distinguish between time orientations:
by convention, one chooses the \emph{future} and \emph{past}
null-cones in $\H$ to be, respectively,
$$\N^+:=\{u\tn\bu,~u\in\U\}~,\quad \N^-:=\{-u\tn\bu,~u\in\U\}~.$$

\begin{proposition}\label{p:existence2spinorPaulibases}
For each $g$-orthonormal positively oriented basis $(\ee_\l)$ of $\H$,
such that $\ee_0$ is timelike and future-oriented,
there exists a normalized 2-spinor basis $(\zeA)$ whose associated
Pauli basis $(\t_\l)$ coincides with $(\ee_\l)$\,.
\end{proposition}

\remark~From the above proposition it follows
that any future-pointing timelike vector can be written as
\hbox{$u\tn\bu+v\tn\bv$}\,,
for suitable $u,v\in\U$\,.
\smallbreak

\subsection{From 2-spinors to 4-spinors}\label{s:From 2-spinors to 4-spinors}

Next observe that an element of $\U\tn\Uc$
can be seen as a linear map $\Ua\to\U$,
while an element of $\Ua\tn\Ul$ can be seen as a linear map $\U\to\Ua$.
Then one defines the linear map
\begin{align*}
&\g:\U\tn\Uc\to\End(\U\oplus\Ua):y\mapsto
\g(y):=\sqrt2\,\bigl(y,y^{\fl\lin}\bigr)~,\phantom{\text{\ie}\quad}\\[6pt]
\text{\ie}\quad
&\g(y)(u,\chi)=\sqrt2\bigl(y\pint\chi\,,u\pint y^\fl \bigr)~,
\end{align*}
where $y^\fl:=g^\fl(y)\in\Ul\tn\Ua$ and $y^{\fl\lin}\in\Ua\tn\Ul$
is the transposed tensor.
In particular for a decomposable $y=p\tn\bq$ one has
$$\td\g(p\tn\bq)(u,\chi)
=\sqrt2\bigl(\bang{\chi,\bq}\,p\,,\bang{p^\fl,u}\,\bq^\fl\,\bigr)~.$$

\begin{proposition}
For all $y,y'\in\U\tn\Uc$ one has
$$\g(y)\comp\g(y')+\g(y')\comp\g(y)=2\,g(y,y')\,\id~.$$
\end{proposition}
\proof
It is sufficient to check the statement's formula for any couple of null
\ie\ decomposable elements in $\U\tn\Uc$.
Using the identity
$$\e(p,q)\,r^\fl+\e(q,r)\,p^\fl+\e(r,p)\,q^\fl=0~,\quad p,q,r\in\U~,$$
which is in turn easily checked, a straightforward calculation gives
\begin{multline*}
[\g(p\tn\bq)\comp\g(r\tn\bs)
+\g(r\tn\bs)\comp\g(p\tn\bq)](u+\chi)=\\[6pt]
=2\,\e(p,r)\,\be(\bq,\bs)\,(u,\chi)=
2\,g(p\tn \bq,r\tn \bs)\,(u,\chi)~.
\end{multline*}
\qed

Now one sees that $\g$ is a \emph{Clifford map} relatively to $g$
(see also \Sec\ref{s:Dirac algebra});
thus one is led to regard $$\W:=\U\oplus\Ua$$ as the space of Dirac spinors,
decomposed into its Weyl subspaces.
Actually, the restriction of $\g$ to the Minkowski space $\H$
turns out to be a Dirac map.

The 4-dimensional complex vector space $\W$ is naturally endowed
with a further structure:
the obvious anti-isomorphism
$$\W\to\Wl:(u,\chi)\mapsto(\bch,\bu)~.$$
Namely, if $\psi=(u,\chi)\in\W$ then $\bar\psi=(\bu,\bch)\in\Wc$
can be identified with $(\bch,\bu)\in\Wl$\,;
this is the so-called `Dirac adjoint' of $\psi$\,.
This operation can be seen as the ``index lowering anti-isomorphism''
related to the Hermitian product
$$\kO:\W\times\W\to\CC:\Bigl((u,\chi),(u',\chi')\Bigr)
\mapsto\bang{\bch,u'}+\bang{\chi',\bu}~,$$
which is obviously non-degenerate;
its signature turns out to be $(+\,+\,-\,-)$,
as it can be seen in a ``Dirac basis'' (below).

Let $(\zeA)$ be a normalized basis of $\U$\,;
the \emph{Weyl basis} of $\W$
is defined to be the basis $(\z_\a)$, $\a=1,2,3,4$, given by
$$(\z_1\,,\z_2\,,\z_3,\z_4):=(\z_1\,,\z_2\,,-\bzz^1,-\bzz^2)~.$$
Above, $\z_1$ is a simplified notation for $(\z_1\,,0)$, and the like.
Another important basis is the \emph{Dirac basis} $(\z'_\a)$, $\a=1,2,3,4$,
where
\begin{align*}
& \z'_1=\osq(\z_1\,,\bzz^1)\equiv\osq(\z_1-\z_3)~,
&& \z'_2=\osq(\z_2\,,\bzz^2)\equiv(\z_2-\z_4)~,\\[6pt]
& \z'_3=\osq(\z_1\,,-\bzz^1)\equiv(\z_1+\z_3)~,
&& \z'_4=\osq(\z_2\,,-\bzz^2)\equiv(\z_2+\z_4)~.
\end{align*}
Setting
$$\g_\l:=\g(\t_\l)\in\End(\W)$$
one recovers the usual Weyl and Dirac representations as the matrices
$\bigl(\g_\l\bigr)$\,, $\l=0,1,2,3$\,,
in the Weyl and Dirac bases respectively.

\subsection{Further structures}\label{s:Further structures}

Some other operations on 4-spinor space, commonly used in the literature,
actually depend on particular choices or conventions.
Similarly to the choice of a basis or of a gauge
they are useful in certain arguments or calculations,
but don't need to be fixed in the theory's foundations.
I'll describe the cases of a Hermitian form on $\U$,
of \emph{charge conjugation}, \emph{parity} and \emph{time reversal};
I'll show the relations among these objects
and how they are related to the notion of \emph{observer}.

A Hermitian 2-form $h$ on $\U$ is an element in $\Ua\vh\Ul$\,,
hence it can be seen as an element in $\H^*$\,;
more precisely, $\bh\in\H^*$\,.
One says that $h$ is \emph{normalized} if it is non-degenerate, positive
and $g^\#(h)=h^{-1}$;
the latter condition is equivalent to $g(h,h)=2$\,.
If $h$ is normalized then it is necessarily
a future-pointing timelike element in $\H^*$\,.
For example, consider the Pauli basis $(\t_\l)$
determined by a normalized 2-spinor basis $(\zeA)$\,,
and let $(\tt^\l)$ be the dual basis;
then $\sqrt2\,\bar\tt^0=\bzz^1\tn\zz^1+\bzz^2\tn\zz^2$ is normalized;
conversely, every positive-definite normalized Hermitian metric $h$
can be expressed in the above form for some suitable
normalized 2-spinor bases.\footnote{
Similarly, negative-definite Hermitian metrics correspond to
past-pointing timelike covectors.
Hermitian metrics of mixed signature $(1,-1)$ correspond to
spacelike covectors;
actually, such metrics can always be written as proportional
to $\sqrt2\,\bar\tt^3=\bzz^1\tn\zz^1-\bzz^2\tn\zz^2$\,,
in appropriate normalized 2-spinor bases.} 

The basic observation resulting from the above discussion is that
the assignments of an `observer' in $\H$ and of a positive-definite
Hermitian metric on $\U$ are equivalent;
actually, the two objects are nearly the same thing.
In 4-spinor terms, the above equivalence is only slightly less obvious.
If $h$ is assigned, then it extends naturally
to a Hermitian metric $h$ on $\W$,
which can be characterized by\footnote{
In the traditional notation,
$\g_\l^\dag$ indicates the $h$-adjoint of $\g_\l$\,,
and then depends on the chosen observer.} 
$$h(\psi,\phi)=\kO(\g_0\psi,\phi)~.$$

Charge conjugation depends on the choice of a normalized $2$-form
$\om=\eO^{\iO t}\,\e\in\weu2\Ul$, and is defined as the anti-isomorphism
$$\Ccal_\om:\W\to\W:\psi\mapsto\Ccal_\om(\psi)\equiv\Ccal(u,\chi)
=\bigl(\om^\#(\bch),-\bar\om^\fl(\bu)\bigr)
=\eO^{-\iO t}\,\bigl(\e^\#(\bch),-\be^\fl(\bu)\bigr)~.$$
Thus $\Ccal_\om=\eO^{-\iO t}\,\Ccal_\e$\,.
One also gets
\begin{align*}&
\Ccal_\om\comp\Ccal_\om=\Id{\sst\W}~,\\&
\g_y\comp\Ccal_\om+\Ccal_\om\comp\g_y=0 \quad\Leftrightarrow \quad
\Ccal_\om\comp\g_y\comp\Ccal_\om=-\g_y~,\quad y\in\H\,.
\end{align*}

Finally, parity is an isomorphism of $\W$
dependent on the choice of an observer,
while time-reversal is an anti-isomorphism dependent on the choice
of an observer \emph{and} of a normalized $2$-form;
they are defined by
$$\Pcal:=\g_0\equiv\g(\t_0)~,\qquad \Tcal_\om:=\gamma_\eta\g_0\Ccal_\om~,$$
where the chosen observer is expressed as $\t_0$ in a suitable Pauli basis,
and $\gamma_\eta$ is the canonical element of the Dirac algebra
corresponding to the $g$-normalized volume form of $\H$,
and expressed in a Pauli basis as $\gamma_\eta=\g_0\g_1\g_2\g_3$
(see~\Sec\ref{s:Dirac algebra}).
\remark~An observer, seen as a Hermitian metric on $\U$,
also determines an isomorphism $\U\tn\Uc\to\U\tn\Ul\equiv\End(\U)$\,.
Through it, one can view `world spinors' as endomorphisms,
thus recovering the algebraic structure for the
Galileian treatment of spin~\cite{CJM}.
\smallbreak

\subsection{2-spinor groups}\label{s:2-spinor groups}

The group $\Aut(\S)\cong\Aut(\U)\subset\U\tn\Ul$ has the natural subgroups
\begin{align*}
& \SlG(\U):=\{K\in\Aut(\U):\det K=1\}~, &&\dim_\CCc\SlG(\U)=3~,
\\[6pt]
& \SlG^c(\U):=\{K\in\Aut(\U):|\det K|=1\}~,&&\dim_\RRr\SlG^c(\U)=7~.
\end{align*}
The former is the group of all automorphisms of $\S$ (of $\U$)
which leave any complex volume form invariant;
the latter is the group of all automorphisms
which leave any complex volume form invariant up to a phase factor,
and thus it can be seen as the group which preserves
the two-spinor structure.
One has the Lie algebras
\begin{align*}
& \Lie\SlG(\U)\cong\{A\in\End(\U):\Tr A=0\}~,\\[6pt]
& \Lie\SlG^c(\U)\cong\{A\in\End(\U):\Re\Tr A=0\}=\iO\,\RR\oplus\Lie\SlG(\U)~.
\end{align*}

If $h\in\Ul\vh\Ua$ is a positive Hermitian metric then one sets
\begin{align*}
&\Ug(\U,h):=\{K\in\Aut(\U):K^\dag=K^{-1}\}\subset\SlG^c(\U)~,\\[6pt]
&\SU(\U,h):=\{K\in\Aut(\U):K^\dag=K^{-1}\,,~\det K=1\}\subset\SlG(\U)~,
\end{align*}
where $K^\dag$ denotes the $h$-adjoint of $K$\,.
One gets the Lie algebras
\begin{align*}
& \Lie\Ug(\U,h)=\{A\in\End(\U):A+A^\dag=0\}=\iO\,\RR\oplus\Lie\SU(\U,h)~,
\\[6pt]
& \Lie\SU(\U,h)=\{A\in\End(\U):A+A^\dag=0\,,~\Tr A=0\}~.
\end{align*}

Now observe that $\End(\U)$ can be decomposed into the direct sum
of the subspaces of all $h$-Hermitian and anti-Hermitian endomorphisms;
the restriction of this decomposition to $\Lie\SlG(\U)$ gives then
$$\Lie\SlG(\U)=\Lie\SU(\U,h)\oplus\iO\,\Lie\SU(\U,h)~.$$

When a 2-spinor basis is fixed, then one gets group isomorphisms
$\SlG(\U)\to\SlG(2,\CC)$\,, $\SU(\U,h)\to\SU(2)$ and the like.

\subsection{2-spinor groups and Lorentz group}
  \label{s:2-spinor groups and Lorentz group}

Up to an obvious transposition we can make the identification
$$\End(\U)\tn\End(\Uc)\cong\End(\U\tn\Uc)~.$$
We then write\footnote{
The elements of the dual Pauli basis can be written as
$\tt^\l=\t\Ii\l\AAd\,\zzA\tn\bzzA$ with
$\t\Ii\l\AAd=g^{\l\m}\,\e^{\sA\sB}\,\be^{\cA\cB}\,\t\iI\m\BBd$\,.
} 
\begin{align*}
&(K\tn\bar H)\Ii{\AAd}{\BBd}=K\Ii\sA\sB\,\bar H\Ii\cA\cB~,
\quad K\in\End(\U)~,\\[6pt]
&(K\tn\bar H)\Ii\l\m=K\Ii\sA\sB\,\bar H\Ii\cA\cB\,\t\Ii\l\AAd\,\t\iI\m\BBd~.
\end{align*}
The group $\Aut(\U)\times\Aut(\Uc)$ can be identified
with the subgroup of $\Aut(\U\tn\Uc)$
constituted of all elements of the type $K\tn\bar H$ with $K,H\in\Aut\U$\,.
This subgroup is sometimes written as $\Aut(\U)\tn\Aut(\Uc)$\,,
which of course must not be intended as a true tensor product.
It has the proper subgroup $\Aut(\U)\vh\Aut(\Uc)$\,,
constituted of all automorphisms of the type $K\tn\bar K$\,, $K\in\Aut(\U)$\,.

\begin{proposition}
$\Aut(\U)\vh\Aut(\Uc)$ preserves the splitting $\U\tn\Uc=\H\oplus\iO\,\H$
and the causal structure of $\H$\,.
\end{proposition}
\proof
There exist bases of $\H$ composed of isotropic elements;
these are also complex bases of isotropic elements of $\U\tn\Uc$.
Then $A\in\Aut(\U\tn\Uc)$ preserves the splitting and the causal structure
iff it sends any element of the form $u\tn\bu$
in an element of the form $v\tn\bv$\,.\qed

Accordingly, on sets
$$\SlG^c(\U)\vh\SlG^c(\Uc)=\SlG(\U)\vh\SlG(\Uc)
:=\{K\tn\bar K:K\in\SlG(\U)\,\}~.$$
Since $K$ preserves $\e$ up to a phase factor,
$K\tn\bar K$ preserves $\e\tn\be\equiv g$\,;
moreover it is immediate to check that any Pauli basis is transformed
to another Pauli basis.
From proposition~\ref{p:existence2spinorPaulibases} it then follows
that $\SlG(\U)\vh\SlG(\Uc)$ restricted to $\H$ coincides
with the special ortochronous Lorentz group $\Lor_+^\up(\H,g)$\,.
Actually, the epimorphism $\SlG(\U)\to\Lor_+^\up(\H,g)$
turns out to be \hbox{2-to-1}\,.

The Lie algebra of $\SlG(\U)\vh\SlG(\Uc)$ is the Lie subalgebra of
$\End(\U)\tn\End(\Uc)$ constituted by all elements
which can be written in the form
$$A\tn\Id{\Uc}+\Id{\U}\tn\bar A~,\quad A\in\Lie\SlG(\U)~.$$
One easily checks that these restrict to endomorphisms of $\H$,
actually they constitute the vector space of all $g$-antisymmetric
endomorphisms of $\H$ namely the Lie algebra $\Lie\Lor(\H,g)$\,.
Let a normalized 2-spinor basis be fixed;
then the isomorphism $\Lie\SlG(\U)\leftrightarrow\Lie\Lor(\H,g)$\,,
taking into account the isomorphism $\Lie\Lor(\H,g)\leftrightarrow\weu2\H^*$
induced by the Lorentz metric $g$\,,
associates the basis $(\n_i\,;\check\n_i)$
with the basis $(\r_i\,;\check\r_i)$\,, $i=1,2,3$\,,
where\footnote{
Here again $(\s\iIi i\sA\sB)$ denotes the $i$-th Pauli matrix.
$(\tt^\l)$ is the dual Pauli basis.
Also note that the Hodge isomorphism restricts to a complex structure
on $\weu2\H^*$.} 
\begin{align*}
& \n_i:=-\iO\,\check\n_i
&&\check\n_i:=\oh\,\s_i\equiv\oh\,\s\iIi i\sA\sB\,\zeA\tn\zzB~,
~,\\[6pt]
&\r_i:=-{*}\check\r_i~, &&\check\r_i:=2\,\tt^0\we\tt^i~.
\end{align*}

A Hermitian metric $h$ on $\U$, besides the above said
(\Sec\ref{s:2-spinor groups})
splitting of $\Lie\SlG(\U)$\,,
also determines an ``observer'' $\t_0:=\osq\,\bh^\#$\,,
hence also the splitting of $\Lie\Lor(\H,g)$ into ``infinitesimal rotations''
and ``infinitesimal boosts'' as
$$\Lie\Lor(\H,g)=\Lie\Lor_{\sst\mathrm{R}}(\H,g,\t_0)
\oplus\Lie\Lor_{\sst\mathrm{B}}(\H,g,\t_0)~.$$
If one chooses a normalized 2-spinor basis
such that the element $\t_0$ of the corresponding Pauli basis of $\H$
coincides with the given observer,
then the bases $(\n_i\,;\check\n_i)$ and $(\r_i\,;\check\r_i)$
turn out to be adapted to the respective splittings.

\remark~On $\Lie\Lor(\H,g)$ one has the pseudo-metric induced by $g$\,;
moreover, consider the real symmetric 2-form
$$\KO_{\sst\Lie\SlG}:\Lie\SlG(\U)\times\Lie\SlG(\U)\to\RR:
(A,B)\mapsto 2\,\Re\Tr(A\comp B)~.$$
Then it turns out that the bases
$(\n_i\,;\check\n_i)$ and $(\r_i\,;\check\r_i)$
are orthonormal,
and that the signature of both metrics is
$(-\,,\,-\,,\,-\,,\,+\,,\,+\,,\,+)$\,.
So, the splittings of the two algebras
determined by the choice of an ``observer''
can't be into arbitrary subspaces:
the two components must be mutually orthogonal subspaces
of opposite signature.

\section{Two-spinor bundles}
\subsection{Two-spinor connections}\label{s:Two-spinor connections}

Consider any real manifold $\M$ and a vector bundle
$\S\to\M$ with complex $2$-dimensional fibres.
Denote base manifold coordinates as $(\xx^a)$;
choose a local frame $(\xi_\sA)$ of $\S$,
determining linear fibre coordinates $(\xx^\sA)$.
According to the constructions of the previous sections,
one now has the bundles $\Q$, $\LL$, $\U$, $\H$ over $\M$,
with smooth natural structures;
the frame $(\xi_\sA)$ yields the frames $\e$, $l$, $(\zeA)$ and $(\t_\l)$\,,
respectively.
Moreover for any rational number $r\in\QQ$
one has the semi-vector bundle $\LL^r$\,.

Consider an arbitrary $\CC$-linear connection $\Cs$ on $\S\to\M$,
called a \emph{$2$-spinor connection}.
In the fibred coordinates $(\xx^a,\xx^\sA)$\, $\Cs$ is expressed
by the coefficients $\Cs\iIi{a}{\sA}{\sB}:\M\to\CC$\,,
namely the covariant derivative of a section $s:\M\to\S$ is expressed as
$$\nabla s=(\de_a s^\sA-\Cs\iIi{a}{\sA}{\sB}s^\sB)\,\dx^a\tn\xi_\sA~.$$
The rule
$\nabla\bs=\overline{\nabla s}$
yields a connection $\bar\Cs$ on $\Sc\to\M$,
whose coefficients are given by
$$\bar\Cs\iIi{a}{\cA}{\cB}=\overline{\Cs\iIi{a}{\sA}{\sB}}~.$$
Actually, $\Cs$ determines linear connections on each of the above said
induced vector bundles over $\M$
(in particular, it is easy to see that any $\CC$-linear connection
on a complex vector bundle determines a $\RR$-linear connection
on the induced Hermitian tensor bundle).
Denote by $2\,G$ and $2\,Y$ the connections induced on $\LL$ and $\Q$
(this notation makes sense because the fibres are 1-dimensional), namely
\begin{gather*}
\nabla l=-2\,G_a\,\dx^a\tn l~,\quad  \nabla\e=2\,\iO\,Y_a\,\dx^a\tn\e~,\\
\nabla\ww^{-1}\equiv\nabla(l^{-1}\tn\e)=2(G_a+\iO\,Y_a)\,\dx^a\tn l^{-1}\tn\e
\end{gather*}
and the like. By direct calculation we find
\begin{align*}
G_a&=\Re(\oh\,\Cs\iIi{a}{\sA}{\sA})=
\oq(\Cs\iIi{a}{\sA}{\sA}+\bar\Cs\iIi{a}{\cA}{\cA})~,
\\[8pt]
Y_a&=\Im(\oh\,\Cs\iIi{a}{\sA}{\sA})=
\tfrac{1}{4\iO}(\Cs\iIi{a}{\sA}{\sA}-\bar\Cs\iIi{a}{\cA}{\cA})~.
\end{align*}
Note that since $Y_a$ are real the induced linear connection on $\Q$
is Hermitian (preserves its natural Hermitian structure).

The coefficients of the connection $\td\Cs$ induced on $\U$ are given by
$$\td\Cs\iIi a\sA\sB=\Cs\iIi a\sA\sB-G_a\,\d\Ii\sA\sB~.$$

Let $\td\G$ be the connection induced on $\U\tn\Uc$,
and $\G'$ the connection induced on $\S\tn\Sc$. Then
\begin{align*}
\G'\iIi{a}{\AAd}{\BBd}&=\Cs\iIi a\sA\sB\,\d\Ii\cA\cB
+\d\Ii\sA\sB\,\bar\Cs\iIi a\cA\cB~,
\\[8pt]
\td\G\iIi{a}{\AAd}{\BBd}&=
\Cs\iIi a\sA\sB\,\d\Ii\cA\cB+\d\Ii\sA\sB\,\bar\Cs\iIi a\cA\cB
	-2\,G_a\,\d\Ii\sA\sB\,\d\Ii\cA\cB~.
\end{align*}
Since the above coefficients are real,
$\G'$ and $\td\G$ turn out to be reducible to real connections on
$\S\vh\Sc$ and $\H\equiv\U\vh\Uc$, respectively.
Moreover
\begin{proposition}
The connection $\td\G$ induced on $\H$ by any $2$-spinor connection is metric,
namely $\nabla[\td\G]g=0$\,.
\end{proposition}
\proof
The Lorentz metric $g$ of $\H$
can be identified with the identity of the bundle $\LL^{-2}$,
namely it is the canonical section
$1\equiv\e^{-1}\tn\e:\M\to\LL^{-2}\tn\LL^2\equiv\M\times\RR^+$\,,
which obviously has vanishing covariant derivative.
\qed

Because of metricity the coefficients $\td\G\iIi a\l\m$ of $\td\G$
in the frame $(\t_\l)$ are antisymmetric and traceless, namely
$$\td\G\iI{a}{\l\m}+\td\G\iI{a}{\m\l}=0~,\quad\td\G\iIi{a}{\l}{\l}=0$$
(the second formula says $\nabla\eta=0$\,,
where $\eta$ is the $g$-normalized volume form of $\H$).

The above relations between $\Cs$ and the induced connections can be inverted
as follows:
\begin{proposition} One has
$$\Cs\iIi a\sA\sB=(-G_a+\iO\,Y_a)\,\d\Ii\sA\sB+\oh\,\G'\iIi a{\AAd}{\sB\cA}
=(G_a+\iO\,Y_a)\,\d\Ii\sA\sB+\oh\,\td\G\iIi a{\AAd}{\sB\cA}~.$$
\end{proposition}

\smallbreak
In $4$-spinor formalism the above relation reads
$$\Cs\iIi{a}{\a}{\b}=
(G_a+\iO\,Y_a)\,\d\Ii{\a}{\b}
+\oq\,\td\G\iI{a}{\l\m}(\g_\l\,\g_\m)\Ii{\a}{\b}~,$$
where now $\Cs\iIi{a}{\a}{\b}$ stands for the coefficients of the
naturally induced connection $(\Cs,\bar\Cs^\lin)$ on $\W\equiv\U\dir{\M}\Ua$
in any 4-spinor frame, $\a,\b=1,..,4$.

A similar relation holds among the curvature tensors, namely
\begin{align*}
R\iIi{ab}{\sA}{\sB}&=
2\,(\dO G-\iO\,\dO Y)_{ab}\,\d\Ii{\sA}{\sB}+\oh\,R'\iIi{ab}{\AAd}{\sB\cA}=
\\[8pt]
&
=-2\,(\dO G+\iO\,\dO Y)_{ab}\,\d\Ii{\sA}{\sB}+\oh\,\td R\iIi{ab}{\AAd}{\sB\cA}~,
\end{align*}
where $R$, $R'$ and $\td R$ are the curvature tensors of
$\Cs$, $\G'$ and $\td\G$, respectively.

\remark~Under a local gauge transformation $\Ksf:\M\to\Gl(2,\CC)$
the above coefficients transform as
\begin{align*}
&\Cs\iIi{a}{\sA}{\sB} \mapsto
(\Ksf^{-1})^\sA_\sC\,\Ksf^\sD_\sB\,\Cs\iIi{a}{\sC}{\sD}
-(\Ksf^{-1})^\sA_\sC\,\de_a\Ksf^\sC_\sB~,
\\[6pt]
&G_a \mapsto G_a-\oh\,\de_a\log\left|\det \Ksf\right|~,\quad
Y_a \mapsto Y_a-\oh\,\de_a\arg\det \Ksf~,
\\[6pt]
& \td\G\iIi{a}{\l}{\m} \mapsto
(\td \Ksf^{-1})^\l_\n\,\td \Ksf^\r_\m\,\td\G\iIi{a}{\n}{\r}
-(\td \Ksf^{-1})^\l_\n\,\de_a\td \Ksf^\n_\m~.
\end{align*}
\smallbreak

\subsection{Two-spinor tetrad}\label{s:Two-spinor tetrad}

Henceforth I'll assume that $\M$ is a real $4$-dimensional manifold.
Consider a linear morphism
$$\Th:\TO\M\to\S\tn\Sc=\CC\tn\LL\tn\H~,$$
namely a section
$$\Th:\M\to\CC\tn\LL\tn\H\tn\TS\M$$
(all tensor products are over $\M$).
Its coordinate expression is
$$ \Th=\Th_a^\l\,\t_\l\tn\dx^a=\Th_a^\AAd\,\zeA\tn\bzeA\tn\dx^a~,
\qquad \Th_a^\l,\Th_a^\AAd:\M\to\CC\tn\LL~.$$

We'll assume that $\Th$ is non-degenerate
and valued in the Hermitian subspace $\LL\tn\H\subset\S\tn\Sc$\,;
then $\Th$ can be viewed as a `scaled' \emph{tetrad}
(or \emph{soldering form}, or \emph{vierbein});
the coefficients $\Th_a^\l$ are real (\ie\ valued in $\RR\tn\LL$)
while the coefficients $\Th_a^\AAd$ are Hermitian,
\ie\ $\bar\Th_a^{\cA\sA}=\Th_a^\AAd$.

\remark~Most of what follows actually still holds
in the case of a degenerate tetrad.
The inverse $\Th^{-1}$ is not used.
This will give rise to a more natural theory,
in which all field equations are of the first order.
Possible degeneracy might also have a physical meaning,
as discussed in~\cite{C98}.
\smallbreak

Through a tetrad, the geometric structure of the fibres of $\H$
is carried to a similar, scaled structure on the fibres of $\TO\M$.
It will then be convenient, from now on, to distinguish by a tilda
the objects defined on $\H$,
so I'll denote by $\td g$\,, $\td\eta$ and $\td\g$
the Lorentz metric,
the $\td g$-normalized volume form and the Dirac map of $\H$\,, and set
\begin{align*}
g&:=\Th^*\td g:\M\to\CC\tn\LL^2\tn\TS\M\tn\TS\M~,
\\[6pt]
\eta&:=\Th^*\td\eta:\M\to\CC\tn\LL^4\tn\weu{4}\TS\M~,
\\[6pt]
\g&:=\td\g\comp\Th:\TO\M\to\LL\tn\End(\W)~,
\end{align*}
which have the coordinate expressions
\begin{align*}
g&=\eta_{\l\m}\,\Th_a^\l\,\Th_b^\m\,\dx^a\tn\dx^b
=\e_{\sA\sB}\e_{\cA\cB}\,\Th_a^\AAd\,\Th_b^\BBd\,\dx^a\tn\dx^b~,
\\[6pt]
\eta&=\det(\Th)\,\dx^0\we\dx^1\we\dx^2\we\dx^3~,
\\[6pt]
\g&=\sqrt2\,\Th_a^\AAd\,
(\zeA\tn\bzeA+\e_{\sA\sB}\e_{\cA\cB}\,\bzzB\tn\zzB)\tn\dx^a~.
\end{align*}
The above objects turn out to be a Lorentz metric,
the corresponding volume form and a Clifford map.
Moreover
$$ \Th_\m^b:=\Th_a^\l\,\eta_{\l\m}\,g^{ab}=(\Th^{-1})_\m^b:\M\to\CC\tn\LL^{-1}~,
\quad
g^{ab}:\M\to\CC\tn\LL^{-2}~.$$

A non-degenerate tetrad, together with a two-spinor frame,
yields mutually dual orthonormal frames
$(\Th_\l)$ of $\LL^{-1}\tn\TO\M$
and $(\ost\Th{}^\l)$ of $\LL\tn\TS\M$\,, given by
$$\Th_\l:=\Th^{-1}(\t_\l)=\Th_\l^a\,\de\xx_a~,\quad
\ost\Th{}^\l:=\Th^*(\tt^\l)=\Th_a^\l\,\dx^a~.$$

We also write
\begin{align*}&
\g=\g_\l\tn\ost\Th{}^\l=\g_a\tn\dx^a~,\quad
\g_\l:=\g(\Th_\l):\M\to\End(\W)~,\\&
\g_a:=\g(\de\xx_a)=\Th_a^\l\,\g_\l:\M\to\LL\tn\End(\W)~.
\end{align*}

\subsection{Cotetrad}

One defines a natural `exterior' product of elements in
the fibres of $\H\ten{\M}\TS\M$
by requiring that, for decomposable tensors, it is given by
$$(y_1\tn\a_1)\we(y_2\tn\a_2)=(y_1\we y_2)\tn(\a_1\we\a_2)~,
\quad \a_1\,,\a_2\in\TS\M\,,~u_1\,,u_2\in\H~.$$
We'll consider the exterior products
$$\weu{q}\Th:\M\to\CC\tn\LL^q\tn\weu{q}\H\tn\weu{q}\TS\M~,
\quad q=1,2,3,4~.$$
In particular, one has $\weu2\Th\equiv\Th\we\Th$\,, that is
$$\weu{2}\Th(u\we v)=\Th(u)\we\Th(v) \qRq
\weu2\Th=\Th_a^\l\Th_b^\m\,(\t_\l\we\t_\m)\tn(\dx^a\we\dx^b)~.$$

Next, consider the linear map over $\M$
$$\sTh:(\S\tn\Sc)\tn\TS\M\to\CC\tn\LL^4\tn\weu{4}\TS\M$$
defined by
$$\sTh(\xi):=\tfrac{1}{3!}\,\td\eta\mid(\xi\we\Th\we\Th\we\Th)=
\tfrac{1}{3!}\,\td\eta\mid[\xi\we(\weu{3}\Th)]~.$$
Its coordinate expression is
\begin{align*}
& \sTh(\xi)=\sTh_\l^a\,\xi_a^\l\,\dO^4\xx:=
\tfrac{1}{3!}\,\e^{abcd}\,\e_{\l\m\n\r}\,
\Th_b^\m\Th_c^\n\Th_d^\r\,\xi_a^\l\,\dO^4\xx~,
\\[6pt]
& \xi=\xi_a^\l\,\t_\l\tn\dx^a~,\quad \xi_a^\l:\M\to\CC\tn\LL~.
\end{align*}

Now $\sTh$ can be seen as a bilinear map
$(\S\tn\Sc)\times\TS\M\to\CC\tn\LL^4\tn\weu{4}\TS\M$
over $\M$, or also as a linear map
$$\S\tn\Sc\to\CC\tn\LL^4\tn\TO\M\tn\weu{4}\TS\M$$
over $\M$.
Using the latter point of view, if $\Th$ is non-degenerate then one has
$$\sTh=\Th^{-1}\tn\eta~.$$
Namely, in general one may regard $\sTh$\,,
which is called the \emph{co-tetrad},
as a kind of `pseudo-inverse' of $\Th$\,,
defined even if $\Th$ is degenerate.

The above construction can be easily generalized, for $p=0,1,2,3,4$,
to a map
$$\sTh^{(p)}:
\weu{p}(\S\tn\Sc)\tn(\weu{p}\TS\M)\to\CC\tn\LL^4\tn\weu{4}\TS\M~.$$
We'll be concerned with $\sTh^{(1)}=\sTh$ and $\sTh^{(2)}$.
Note that $\sTh^{(0)}=\eta$.

\subsection{Tetrad and connections}\label{s:Tetrad and connections}

If $\Cs$ is a complex-linear connection on $\S$,
and $G$ and $\td\G$ are the induced connections on $\LL$ and $\H$,
then a non-degenerate tetrad $\Th:\TO\M\to\LL\tn\H$ yields a unique
connection $\G$ on $\TO\M$, characterized by the condition
$$\nabla[\G\tn\td\G]\Th=0~.$$
Moreover $\G$ is metric, \ie\ $\nabla[\G]g=0$.
Denoting by $\G\iIi{a}{\l}{\m}$ the coefficients of $\G$
in the frame $\Th_\l'\equiv\Th^{-1}(l\tn\t_\l)$ one obtains
$$\G\iIi{a}{\l}{\m}=\td\G\iIi{a}{\l}{\m}+2\,G_a\,\d\Ii{\l}{\m}~.$$

The curvature tensors of $\G$ and $\td\G$ are related by
$R\iIi{ab}{\l}{\m}=\td R\iIi{ab}{\l}{\m}$\,, or
$$R\iIi{ab}{c}{d}=\td R\iIi{ab}{\l}{\m}\,\Th_\l^c\,\Th_d^\m~.$$
Hence the Ricci tensor and the scalar curvature are given by
\begin{align*}
R_{ad}&=R\iIi{ab}{b}{d}=\td R\iIi{ab}{\l}{\m}\,\Th_\l^b\,\Th_d^\m~,\\
R\iI{a}{a}&=\td R\iI{ab}{\l\m}\,\Th_\l^b\,\Th_\m^a~.
\end{align*}

In general,
the connection $\G$ will have non-vanishing torsion,\footnote{ 
This is the tensor field $T:\M\to\TO\M\tn\weu2\TS\M$ defined by
$T(u,v)=\na_uv-\na_vu-[u,v]$\,,
where $u,v:\M\to\TO\M$ are any two vector fields,
and has the coordinate expression
$T\Ii c{ab}=-\G\iIi acb+\G\iIi bca$\,.} 
which can be expressed\footnote{ 
Taking into account
$0=\na_a\Th_b^\l=\de_a\Th_b^\l-\G\iIi a\l\m\,\Th_b^\m+\G\iIi acb\,\Th_c^\l$\,.} 
as
$$\Th_c^\l\,T\Ii{c}{ab}=\de_{[a}^\phexp\Th_{b]}^\l
+\Th_{[a}^\m\,\td\G\iIi{b]}{\l}{\m}
+2\,\Th_{[a}^\l\,G^\phexp_{b]}~.$$

\remark~The torsion can be seen as the Fr\"olicher-Nijenhuis bracket
$$\td T:=T\pint\Th=[\G',\Th]:\M\to\weu2\TS\M\ten{\M}H'~,$$
where $H'=\LL\tn\H$,
$\G':\H'\to\TS\M\ten{\H'}\TO\H'$
is the induced connection on $\H'\to\M$,
and $\Th$ is seen as a vertical-valued form
$\Th:\H'\to\TS\M\ten{\H'}\VO\H'$\,.
\smallbreak

\subsection{The Dirac operator}\label{s:The Dirac operator}

Given a tetrad and a two-spinor connection,
one introduces the Dirac operator acting on sections
$\psi:\M\to\LL^{-3/2}\tn\W$.

Writing
$\td\g^\#:\M\to\H\tn\End(\W)$\,,
$\nabla\psi:\M\to\LL^{-3/2}\tn\TS\M\ten{\M}\W$\,,
one has
$$\td\g^\#\nabla\psi:\M\to\LL^{-3/2}\tn\H\tn\TS\M\tn\W~,$$
where contraction in $\W$ is understood.
Next, one contracts the factors $\H$ and $\TS\M$ above via
$$\sTh:\M\to\CC\tn\LL^3\tn\H^*\tn\TO\M\tn\weu4\TS\M~,$$
obtaining
$$\breve\nasl\psi:=\bang{\sTh,\td\g^\#\nabla\psi}:
\M\to\LL^{3/2}\tn\W\tn\weu4\TS\M~,$$
which has the coordinate expression
$$\breve\nasl\psi=\sTh^a_\l\,\left(
\s^{\l\AAd}\,\na_a\chi_\cA\,\zeA\,,\,\s\Ii{\l}{\AAd}\na_a u^\sA\,\bzzA\,
\right)\tn\dO^4\xx~.$$
This definition works even if $\Th$ were degenerate;
in the non-degenerate case one simply has $\breve\nasl\psi=\nasl\psi\tn\eta$\,.

\section{Two-spinors and field theories}
\label{S:Two-spinors and field theories}
\subsection{The fields}\label{s:The fields}

In this section I'll present a ``minimal geometric data'' field theory:
actually, the unique ``geometric datum'' is a vector bundle $\S\to\M$
with complex 2-dimensional fibres and real 4-dimensional base manifold.
All other bundles and fixed geometric objects are determined just
by this datum through functorial constructions,
as we saw in the previous sections;
no further background structure is assumed.
Any considered bundle section which is not functorially fixed
by our geometric datum is a field.
In this way one obtains a field theory
which turns out to be essentially equivalent to a classical theory
of Einstein-Cartan-Maxwell-Dirac fields.

The fields are taken to be the tetrad $\Th$\,,
the $2$-spinor connection $\Cs$,
the electromagnetic field $F$ and the electron field $\psi$\,.
The gravitational field is represented by $\Th$
(which can be viewed as a `square root' of the metric)
and the traceless part of $\Cs$, namely $\td\G$,
seen as the gravitational part of the connection.
If $\Th$ is non-degenerate one obtains,
as in the standard metric-affine approach~\cite{GrHe,HCMN,Re,FK82},
essentially the Einstein equation and the equation for torsion;
the metricity of the spacetime connection is a further consequence.
But note that the theory is non-singular also in the degenerate case.
The connection $G$ induced on $\LL$ will be assumed to have vanishing curvature,
$\dO G=0$, so that one can always find local charts such that $G_a=0$;
this amounts to gauging away the conformal (`dilaton') symmetry.
Coupling constants will arise as covariantly constant sections of $\LL$,
which now becomes just a vector space.

The Dirac field is a section
$$\psi:\M\to\LL^{-3/2}\tn\W:=\LL^{-3/2}\tn(\U\oplus\Ua)~,$$
assumed to represent a semiclassical particle with one-half spin,
mass $m\in\LL^{-1}$ and charge $q\in\RR$\,.

The electromagnetic potential can be thought of as
the Hermitian connection $Y$ on $\weu2\U$ determined by $\Cs$\,,
whose coefficients are indicated as $\iO\,Y_a$\,;
locally one writes
$$Y_a\equiv q\,A_a~,$$
where $A:\M\to\TS\M$ is a local 1-form.

The electromagnetic field is represented by a spinor field
$$\td F:\M\to\LL^{-2}\tn\weu2\H^*$$
which, via $\Th$\,, determines the 2-form $F:=\Th^*\td F:\M\to\weu2\TS\M$\,.
The relation between $Y$ and $F$ will follow as one of the field equations;
note how this setting allows a first-order linear Lagrangian and
non-singularity in the degenerate case also for the electromagnetic sector.

The total Lagrangian and the Euler-Lagrange operator will be the sum of
a gravitational, an electromagnetic and a Dirac term
$$\Lcal=\Lcal\grav+\Lcal\emag+\Lcal\Dir~,\quad
\Ecal=\Ecal\!\grav+\Ecal\!\emag+\Ecal\!\Dir~.$$
Observe that all Lagrangian $4$-forms are defined
in terms of the cotetrad $\sTh$,
while a direct translation of the standard formulation in terms of our fields
would force one to use $\Th^{-1}$,
resulting in a less simple and natural theory.

\subsection{Gravitational Lagrangian}\label{s:Gravitational Lagrangian}

The tetrad $\Th$ and the curvature tensor $\td R$ of $\td\G$ can be assembled
into a $4$-form $\Lcal\grav$ which, in the non-degenerate case,
turns out to be the usual gravitational Lagrangian density:
$$\Lcal\grav:=\frac{1}{4\,\Bbbk}\,\sTh^{(2)}(\td R^\#)=
\frac{1}{8\,\Bbbk}\,\td\eta\mid(\td R^\#\we\Th\we\Th):\M\to\weu4\TO^*\M~,$$
where $\td R^\#:\M\to\weu{2}\TO^*\M\tn\weu{2}\H$ is the curvature tensor
of $\td\G$ with one index raised via $\td g$\,,
and $\Bbbk\in\LL^2$ is Newton's gravitational constant.
Note how this is necessary in order to obtain a true (non-scaled)
$4$-form on $\M$ and the correct coupling with the spinor field.
One has the coordinate expression
$\Lcal\grav=\ell\grav\,\dO^4\xx$ with
$$\ell\grav=\frac{1}{8\,\Bbbk}\,
\e_{\l\m\n\r}\,\e^{abcd}\,\td R\iI{ab}{\l\m}\,\Th_c^\n\,\Th_d^\r
=\frac{1}{2\,\Bbbk}\,R\,\det\Th~,$$
where $R$ is the scalar curvature
and the last equality holds if $\Th$ is non-degenerate.

A calculation gives the $\Th$- and $\td\G$-components of the gravitational part $\Ecal\!\grav$ of the Euler-Lagrange operator:
\begin{align*}
(\Ecal\!\grav)_\n^c&=
\frac1{4\,\Bbbk}\,\e_{\l\m\n\r}\,\e^{abcd}\,R\iI{ab}{\l\m}\,\th_d^\r~,
\\[6pt]
(\Ecal\!\grav)\Ii{a}{\l\m}&=
\frac1{2\,\Bbbk}\,\e_{\l\m\n\r}\,\e^{abcd}\,
(\de_b\Th_c^\n+\Th_b^\s\,\td\G\iIi{c}{\n}{\s}\,)\,\Th_d^\r~.
\end{align*}
In the non-degenerate case these are essentially the Einstein tensor
and the torsion of the spacetime connection, respectively.
The first, in particular, can be written
$$(\Ecal\!\grav)_\n^c=
\frac1{4\,\Bbbk}\,\Th_{[\l}^a\,\Th_\m^b\,\Th_{\n]}^c\,\det\Th
=\frac1\Bbbk\,(R\iI{ab}{bc}-\oh\,R\iI{db}{bd}\,\d_a^c)\Th_\n^a\,\det\Th~.$$

The $\td\G$-component of $\Ecal\!\grav$ can be expressed
in terms of the torsion as
$$(\Ecal\!\grav)\Ii{a}{\l\m}=\frac1{4\,\Bbbk}\,
\e_{\l\m\n\r}\,\e^{abcd}\,T\Ii{e}{bc}\,\Th_e^\n\,\Th_d^\r~.$$

\subsection{Electromagnetic Lagrangian}\label{s:Electromagnetic Lagrangian}

The electromagnetic potential and the
Maxwell field will be considered independent fields.
The former is represented by a local section
$A:\M\to\TS\M$\,,
related to the connection $Y$ induced by $\Cs$ on $\weu{2}\U$ 
by the relation $Y=q\,A~.$
The Maxwell field is a section
$\td F:\M\to\LL^{-2}\tn\weu2\H^*$,
written in coordinates as
$\td F=\td F_{\l\m}\,\tt^\l\tn\tt^\m$\,.
The e.m.\ Lagrangian density is defined to be
$$\Lcal\emag=\ell\emag\,\dO^4\xx=\Bigl[
-\oh\,\Th^{(2)}(\dO A\tn\td F)+\oq\,(\td F{\cdot}\td F)\Bigr]\,\eta~,$$
with coordinate expression
$$\ell\emag=
-\oq\,\e^{abcd}\,\e_{\l\m\n\r}\,\de_aA_b\,\td F^{\l\m}\,\Th_c^\n\Th_d^\r
+\oq\,\td F^{\a\b}\td F_{\a\b}\,\det\Th~.$$
In the non-degenerate case,
this turns out to be essentially the Lagrangian used in the ADM formalism.

Since $\td F$ does not appear in the other terms of the total Lagrangian,
the $\td F$-component of the field equations is immediately seen to yield
$$-\oh\,\e^{abcd}\,\e_{\l\m\n\r}\,\de_aA_b\,\Th_c^\n\,\Th_d^\r
+\td F_{\l\m}\,\det\Th=0~,$$
which in the non-degenerate case gives
$$F:=\Th^*\td F=2\,\dO A
\quad\Rightarrow\quad
\Lcal\emag=-\oq\,F^2\,\eta~.$$

The $A$-component of the Euler-Lagrange operator is
\begin{equation*}\begin{split}
(\Ecal\!\emag)^a&=\oh\,\e^{abcd}\,\e_{\l\m\n\r}\,
(\de_b\td F^{\l\m}\,\Th_c^\n\,\Th_d^\r+
2\,\td F^{\l\m}\,\de_b\Th_c^\n\,\Th_d^\r)=\\
&=\oh\,\e^{abcd}\,(\dO{*}F)_{bcd}~.
\end{split}\end{equation*}

The $\Th$-component is
$$(\Ecal\!\emag)_\n^c=
-\oh\,\e^{abcd}\,\e_{\l\m\n\r}\,\de_aA_b\,\td F^{\l\m}\Th_d^\r
+\oq\,\td F^2\,\sTh_\n^c~,$$
which in the non-degenerate case becomes
essentially the usual Maxwell stress-energy tensor
$$(\Ecal\!\emag)_\n^c=(F_{ab}F^{ac}-\oq\,F^2\,\d_b^c)\sTh_\n^b~.$$

\subsection{Dirac Lagrangian}\label{s:Dirac Lagrangian}

The Dirac spinor field and its `Dirac adjoint' are sections
\begin{align*}
\psi&=(u,\chi):\M\to\LL^{-3/2}\tn\W=\LL^{-3/2}\tn(\U\oplus\Ua)~,\\
\bar\psi&=(\bch,\bu):\M\to\LL^{-3/2}\tn(\Ul\oplus\Uc)=\LL^{-3/2}\tn\Wl~.
\end{align*}
In coordinates:
\begin{align*}
& u=u^\sA\,\zeA~,\quad \chi=\chi_\cA\,\bzzA~,\quad
u^\sA\,,\chi_\cA\,:\M\to\CC\tn\LL^{-3/2}
\\[6pt]
& \bang{\bar\psi,\psi}=(\bu^\cA\,\chi_\cA+\bch_\sA\,u^\sA):
\M\to\CC\tn\LL^{-3}~.
\end{align*}
The Dirac operator (\Sec\ref{s:The Dirac operator}) yields a section
$$\breve\nasl\psi:\M\to\LL^{3/2}\tn\W\tn\weu4\TO^*\M~,$$
so that
$$\bang{\bar\psi,\breve\nasl\psi}:\M\to\CC\tn\weu4\TO^*\M~.$$

Now we introduce the scalar density
$$\Lcal\Dir=
\ih\,\bigl(\bang{\bar\psi,\breve\nasl\psi}-\bang{\breve\nasl\bar\psi,\psi}\bigr)
-m\,\bang{\bar\psi\,,\psi}\,\eta
:\M\to\weu4\TO^*\M~,$$
where $\breve\nasl\bar\psi:=\overline{\breve\nasl\psi}$\,,
and $m\in\LL^{-1}$ is the described particle's mass.
This is a version of the Dirac Lagrangian which remains non-singular
when $\Th$ is degenerate.
In the non-degenerate case one also has
$$\Lcal\Dir=\bigl[
\ih\,\left(\bang{\bar\psi,\nasl\psi}-\bang{\nasl\bar\psi,\psi}\right)
-m\,\bang{\bar\psi\,,\psi}\bigr]\,\eta~;$$
in $2$-spinor terms this reads
$$\Lcal\Dir=\isq\sTh\Bigl(
\na u\tn\bu-u\tn\na\bu+\td g^\#(\bch\tn\na\chi-\na\bch\tn\chi)\Bigr)
-m\,\Bigl(\bang{\chi,\bu}+\bang{\bch,u}\Bigr)\,\eta~,$$
with the coordinate expression
\begin{multline*}
\ell\Dir=\isq\,\sTh^a_{\AAd}\,\Bigl(\na_au^\sA\,\bu^\cA-u^\sA\,\na_a\bu^\cA
+\e^{\sA\sB}\be^{\cA\cB}(\bch_\sB\,\na_a\chi_\cB-\na_a\bch_\sB\,\chi_\cB\,)
\Bigr)
\\
-m\,(\bch_\sA u^\sA+\chi_\cA\,\bu^\cA\,)\,\det\Th~.
\end{multline*}

Next we compute the Euler-Lagrange operator $\Ecal\!\Dir$\,.
The $\bu$-component is
$$(\Ecal\!\Dir)_\cA=\sqrt2\,\iO\,\sTh^a_{\AAd}\,\na_au^\sA
-m\,\chi_\cA\,\det\Th+\isq\,T_\AAd\,u^\sA~,$$
where
$T_\AAd:=\sTh^a_\AAd\,T\Ii{b}{ab}$
is used for replacing the term with $\de_a\Th_b^\m$
(see \Sec\ref{s:Tetrad and connections}).

The $\bch$-component is
$$(\Ecal\!\Dir)^\sA=
\sqrt2\,\iO\,\sTh^{a\AAd}\,\na_a\chi_\cA-m\,u^\sA\,\det\Th
+\isq\,T^\AAd\,\chi_\cA~,$$
with $\sTh^{a\AAd}:=\sTh^a_{\BBd}\,\e^{\sB\sA}\be^{\cB\cA}$ and
$T^\AAd:=\e^{\sB\sA}\be^{\cB\cA}\,T_\BBd$\,.

The $\td\G$-component is
\begin{multline*}
(\Ecal\!\Dir)^a_{\l\m}=
\tfrac{\iO}{4\,\surd2}\,[
(\sTh^a_{\sA\cC}\,\ti{[\l}{\sD\cC}\t_{\m]\sD\cA}^\phexp
-\sTh^a_{\sC\cA}\,\ti{[\l}{\sC\cD}\t_{\m]\sA\cD}^\phexp)u^\sA\bu^\cA
\\
+(\sTh^{a\sB\cC}\ti{[\l}{\sD\cB}\t_{\m]\sD\cC}^\phexp
-\sTh^{a\sC\cB}\ti{[\l}{\sB\cD}\t_{\m]\sC\cD}^\phexp)
\bch_\sB\chi_\cB\,]~.
\end{multline*}

The $\Th$-component is
\begin{align*}
(\Ecal\!\Dir)^c_\n&=
\e^{abcd}\,\e_{\l\m\n\r}\,\Th_b^\m\,\Th_d^\r \Bigl[
\\ &\qquad
\tfrac{\iO}{2\,\surd2}\,\Bigl(
\na_au^\sA\,\bu^\cA-u^\sA\,\na_a\bu^\cA
+\e^{\sB\sA}\be^{\cB\cA}(\bch_\sB\na_a\chi_\cB-\na_a\bch_\sB\chi_\cB\,)\Bigr)
\t\Ii{\l}{\AAd}
\\ & \hspace{5.5cm}
-\tfrac{1}{3!}\,m\,(\bch_\sA\,u^\sA+\chi_\cA\,\bu^\cA\,)\,
\Th_a^\l\Bigr]=
\\[8pt]
&=
\iq\,\e^{abcd}\,\e_{\l\m\n\r}\,\Th_b^\m\,\Th_d^\r
\Bigl(\bar\psi\td\g^\l\na_a\psi-\Bar{\td\g}^\l\na_a\bar\psi\,\psi\Bigr)
-m\,\bar\psi\psi\,\sTh_\n^c~.
\end{align*}

The $A$-component is simply
$$(\Ecal\!\Dir)^a=
\sqrt2\,q\,\sTh_{\AAd}^a\,\Bigl(
u^\sA\,\bu^\cA+\e^{\sB\sA}\,\be^{\cB\cA}\,\bch_\sB\,\chi_\cB\Bigr)=
q\,\sTh_\l^a\,(\bar\psi\td\g^\l\psi)~.$$

\subsection{Field equations}\label{s:Field equations}

Having calculated the various pieces of
$\Ecal=\Ecal\!\grav+\Ecal\!\emag+\Ecal\!\Dir$,
writing down the field equations $\Ecal=0$ is a simple matter.
These equations are non-singular also when $\Th$ is degenerate;
in the non-degenerate case one expects this approach to reproduce
essentially the usual Einstein-Cartan-Maxwell-Dirac field equations.

The $\Th$-component
$$(\Ecal\!\grav)^c_\n=-(\Ecal\!\emag+\Ecal\!\Dir)^c_\n~,$$
corresponds to the Einstein equation;
actually, as already discussed, in the non-degenerate case
the left-hand side is essentially the Einstein tensor,
while the right-hand side can be viewed as the sum of the energy-momentum
tensors of the electromagnetic field and of the Dirac field.

The $\td\G$-component gives the equation for torsion
$$(\Ecal\!\grav)\Ii{a}{\l\m}=-(\Ecal\!\Dir)\Ii{a}{\l\m}~.$$
From this one sees that the spinor field is a source for torsion,
and that in this context one cannot formulate a torsion-free theory.

It was already seen (\Sec\ref{s:Electromagnetic Lagrangian})
that the $\td F$-component reads $F=2\,\dO A$ in the non-degenerate case,
and of course this yields the first Maxwell equation $\dO F=0$.
The $A$-component is
$$-\oh\,\,\e^{abcd}(\dO{*}F)_{bcd}
+q\,\sTh_\l^a\,(\bar\psi\td\g^\l\psi)=0
\quad\text{\ie}\quad
\oh\,c\,\e^{abcd}(\dO{*}F)_{bcd}=q\,\sTh_\l^a\,(\bar\psi\td\g^\l\psi)~.$$
In the non-degenerate case this gives the second Maxwell equation
$$\oh\,{*}\dO{*}F=j~,$$
where $j:\M\to\tn\TS\M$ is the \emph{Dirac current},
with coordinate expression
$$j:=\frac qc\,\Th_a^\l\,(\bar\psi\td\g_\l\psi)\,\dx^a~.$$

The $\bu$- and $\bch$-components
$(\Ecal\!\Dir)_\sA=0$ and $(\Ecal\!\Dir)^\cB=0$
give the following generalized form of the standard \emph{Dirac equation}:
\begin{equation*}\begin{cases}
\sqrt2\,\iO\,\sTh^a_{\AAd}\,\na_au^\sA-m\,\chi_\cA\,\det\Th
+\isq\,T_\AAd\,u^\sA=0
\\[8pt]
\sqrt2\,\iO\,\sTh^{a\AAd}\,\na_a\chi_\cA-m\,u^\sA\,\det\Th
+\isq\,T^\AAd\,\chi_\cA=0
\end{cases}\quad.\end{equation*}
Denoting by $\breve T$ the $1$-form obtained from the torsion by contraction,
with coordinate expression $\breve T_a\equiv T_a=T\Ii{b}{ab}$\,,
the above equation can be written in coordinate-free form as
$$\left(\iO\,\nasl-m+\ih\,\g^\#(\breve T)\right)\psi=0~.$$

\section{Dirac algebra in two-spinor terms}
\label{S:Dirac algebra in two-spinor terms}
\subsection{Dirac algebra}\label{s:Dirac algebra}

If $\V$ is a finite-dimensional real vector space
endowed with a non-degenerate scalar product,
then its \emph{Clifford algebra} $\C(\V)$
is the associative algebra generated by $\V$
where the product of any $u,v\in\V$ is subjected to the condition
$$u\,v+v\,u=2\,u\cdot v~, \quad u,v\in\V~.$$
The Clifford algebra fulfills the following \emph{universal property}:
if $\A$ is an associative algebra with unity and $\g:\V\to\A$
is a linear map such that
$\g(v)\,\g(v)=v\cdot v~\forall\:v\in\V$\,,
then $\g$ extends to a unique homomorphism $\hat\g:\C(\V)\to\A$\,.
It turns out that $\C(\V)$ is isomorphic, as a vector space,
to the vector space underlying the exterior algebra $\wedge\V$\,;
through this isomorphism one identifies $v_1\we \dots\we v_p$
with the antisymmetrized Clifford product
$$\frac1{p!}\,\bigl(
v_1 v_2{\cdot}{\cdot}v_p-v_2 v_1{\cdot}{\cdot}v_p+\,\cdots~\bigr)$$
where the sum is extended to all permutations of the set $\{1,\dots,p\}$\,,
with the appropriate signs.
In other terms,
one has two distinct algebras on the same underlying vector space:
any element of $\C(\V)$ can be uniquely expressed as a sum of terms,
each of well-defined exterior degree.
For example, one has $u\,v=u\we v+u\cdot v$\,;
from this one sees that the Clifford algebra product does not preserve
the exterior algebra degree, but only its parity:
$\C(\V)$ is $\ZZ_2$-graded.
If $\phi\in\weu{r}\V$, $\th\in\weu{s}\V$,
then the Clifford product $\phi\,\th$ turns out to be a sum of terms of
exterior degree $r{+}s,~r{+}s{-}2~,\dots,|r{-}s|$.

\bigbreak
The Clifford algebra $\D:=\C(\H)$ of Minkowski space $\H$
(\Sec\ref{s:2-spinors and Minkowski space})
is called the \emph{Dirac algebra}.
The Dirac map $\g:\H\to\End(\W)$ is a Clifford map,
hence by virtue of the above said universal property
one can see the Dirac algebra
as a real vector subspace $\D\subset\End(\W)$ of dimension $2^4=16$\,.
Since this coincides with the \emph{complex} dimension
of $\End(\W)\equiv\W\tn\Wl$, one gets $\End(\W)=\CC\tn\D$\,.

The Dirac algebra $\D$ is multiplicatively generated
by $\g(\H)\subset\End(\W)$\,, simply identified with $\H$.
One has the natural decompositions
$$\D=\D^\pmap\oplus\D^\mmap=
\bigl(\RR\oplus\weu2\H\oplus\weu4\H\bigr)\oplus\bigl(\H\oplus\weu3\H\bigr)~,$$
where $\D^\pmap$ and $\D^\mmap$
denote the even-degree and odd-degree subspaces, respectively
(the former is a subalgebra).
Also, one has the distinguished elements
$$1\equiv\Id{\W}\subset\RR\subset\D^\pmap~,\quad
\eta^\#\subset\weu4\H\subset\D^\pmap~,$$
where $\eta^\#\equiv g^\#(\eta)$ is the contravariant tensor corresponding
to the unimodular volume form $\eta$\,.
One gets
$$\eta^\#\,\eta^\#=-1~,\quad
\vth\,\eta^\#={*}\vth~~\forall\vth\in\wedge\H~,$$
where ${*}$ is the Hodge isomorphism.

\subsection{Decomposition of $\End\W$ and $\e$-transposition}
\label{s:Decomposition of EndW and e-transposition}

One has the natural decomposition
$$\End(\W)\equiv\End(\U\oplus\Ua)=
(\U\tn\Ul)\oplus(\U\tn\Uc)\oplus(\Ua\tn\Ul)\oplus(\Ua\tn\Uc)~.$$
Accordingly, any $\Phi\in\End(\W)$ is a 4-uple of tensors,
which will be conveniently written in matricial form as
$$\Phi=\begin{pmatrix}K&P\\Q&J\end{pmatrix}~,\qquad
K\in\U\tn\Ul\,,~P\in\U\tn\Uc\,,~Q\in\Ua\tn\Ul\,,~J\in\Ua\tn\Uc\,.$$

We now introduce an operation which acts on each
of the above 4 types of tensors in a similar way.
This operation, called \emph{$\e$-transposition},
is actually independent of the particular normalized $\e\in\weu2\Ul$ chosen;
it is defined by
\begin{align*}
& \U\tn\Ul\to\Ul\tn\U:K\mapsto\td K:=\bang{\e^\fl\tn\e^\#, K}=
\e_{\sC\sA}\,K\Ii\sC\sD\,\e^{\sD\sB}\,\zzA\tn\zeB~,
\\[6pt]
& \U\tn\Uc\to\Ul\tn\Ua:P\mapsto\td P:=\bang{\e^\fl\tn\be^\fl, P}=
\e_{\sC\sA}\,P^{\sC\cD}\,\be_{\cD\cB}\,\zzA\tn\bzzB~,
\\[6pt]
& \Ua\tn\Ul\to\Uc\tn\U:Q\mapsto\td Q:=\bang{\be^\#\tn\e^\#, Q}=
\be^{\cC\cA}\,Q_{\cC\sD}\,\e^{\sD\sB}\,\zeA\tn\bzeB~,
\\[6pt]
& \Ua\tn\Uc\to\Uc\tn\Ua:J\mapsto\td J:=\bang{\be^\#\tn\be^\fl, J}=
\be^{\cC\cA}\,J\iI\cC\cD\,\be_{\cD\cB}\,\bzeA\tn\bzzB~.
\end{align*}
Namely, \emph{$\e$-transposition} changes the position (either high or low)
of both indices of the tensor it acts on.
For elements in $\U\tn\Uc$ or $\Ua\tn\Ul$ it essentially amounts
to index lowering (resp.\ raising) by the Lorentz metric $g$
in complexified Minkowski space;
for invertible elements
in $\U\tn\Ul\equiv\End(\U)$ or $\Ua\tn\Uc\equiv\End(\Ua)$,
$\e$-transposition amounts to
$$\td X=(\det X)\,(X^{-1})^\lin~,$$
where the superscript $\lin$ denotes standard transposition.

It is clear that $\e$-transposition can be similarly defined\footnote{
One could introduce $\e$-transposition on further spaces
such as $\U\tn\U$, $\U\tn\Ua$ and so on.
These extensions however would depend from the chosen normalized $\e$\,;
phase factors cancel out only in the considered cases.} 
on $\Ul\tn\U$, $\Ul\tn\Ua$, $\Uc\tn\U$ and $\Uc\tn\Ua$,
and in all cases one gets
$$\Tilde{\Tilde X}=X~,\quad (\td X)^\lin=(X^\lin)^{\sst\sim}~,\quad
\td X\,X^\lin=X^\lin \td X=(\det X)\,\id~,\quad \det X=\det\td X~.$$
\begin{remark}
The determinant is uniquely defined, via any $\e$\,,
also for elements in $\U\tn\Uc$, $\Ul\tn\Ua$, $\Uc\tn\U$ and $\Ua\tn\Ul$.
In these cases, the determinant of a tensor
equals one-half its Lorentz pseudo-norm.
\end{remark}\smallbreak

Moreover, whenever the composition of tensors $X$ and $Y$ is defined, one has
$$(X\,Y)^{\sst\sim}=\td X\,\td Y~,\quad
\Tr(\td X\,\td Y)=\Tr(X\,Y)~.$$
Whenever $A$ and $B$ are tensors of the same type, one has
$$\det(A+B)=\det(A)+\det(B)+\Tr(A^\lin\td B)~,$$
where the \emph{scalar product} $(A,B)\mapsto\Tr(A^\lin\td B)$
is \emph{symmetric}.\footnote{
On $\U\tn\Uc$ and $\Uc\tn\U$
(resp.\ $\Ul\tn\Ua$  and $\Ua\tn\Ul$)
this coincides with $2\,g$ (resp.\ $2\,g^\#$).} 

\begin{proposition}\label{p:inverseofPhi}
Let $\Phi=\begin{pmatrix}K&P\\ Q&J\end{pmatrix}\in\W\tn\Wl$ be non-singular.
Then
\begin{align*}
& \det\Phi=(\det K)\,(\det J)+(\det P)\,(\det Q)
-\Tr(K^\lin\,\td P\,J^\lin\,\td Q)~,
\\[8pt]
& (\det\Phi)\,\Phi^{-1}=\begin{pmatrix}
(\det J)\,\td K{}^\lin-\td Q{}^\lin\,J\,\td P{}^\lin &&
(\det P)\,\td Q{}^\lin-\td K{}^\lin\,P\,\td J{}^\lin \\[6pt]
(\det Q)\,\td P{}^\lin-\td J{}^\lin\,Q\,\td K{}^\lin &&
(\det K)\,\td J{}^\lin-\td P{}^\lin\,K\,\td Q{}^\lin
\end{pmatrix}~.
\end{align*}
\end{proposition}
\proof It can be checked by a direct calculation,
taking into account the above identities.\qed

\subsection{$\e$-adjoint and characterization of $\D$}
\label{s:e-adjoint and characterization of D}

If $X$ is a tensor of any of the above types, then its \emph{$\e$-adjoint}
is the tensor
$$X^\ddag:=\tb X~.$$
Using this operation one defines the real involution
$$\ddag:\W\tn\Wl\to\W\tn\Wl:
\begin{pmatrix}K & P \\ Q & J\end{pmatrix}\mapsto
\begin{pmatrix}J^\ddag & Q^\ddag \\ P^\ddag & K^\ddag \end{pmatrix}~.$$

\begin{proposition}
$\D$ and $\iO\,\D$ are the eigenspaces of $\ddag$
corresponding to eigenvalues $+1$ and $-1$\,, respectively.
Namely, $\D$ is the real subspace of $\W\tn\Wl$ constituted by
all endomorphisms which can be written in the form
$$\begin{pmatrix}K & P \\ P^\ddag & K^\ddag \end{pmatrix}~,\qquad
K\in\U\tn\Ul~,~~P\in\U\tn\Uc~.$$
Moreover one has the following characterisations
\begin{align*}
&\D^0\equiv\RR=\Bigl\{r\begin{pmatrix}\Id{\U} & 0 \\ 0& \Id{\Ua}\end{pmatrix}~,
\quad r\in\RR\Bigr\}~,
\dbk[8pt]
&\D^1\equiv\H=\Bigl\{\begin{pmatrix}0 & P \\ P^\ddag & 0\end{pmatrix}~,
\quad P\in\H\Bigr\}~,
\dbk[8pt]
&\D^2\equiv\weu2\H=\Bigl\{\begin{pmatrix}K & 0 \\ 0 & K^\ddag\end{pmatrix}~,
\quad K\in\U\tn\Ul\,,~\Tr K=0\Bigr\}~,
\dbk[8pt]
&\D^3\equiv\weu3\H=\Bigl\{\begin{pmatrix}0 & P \\ P^\ddag & 0\end{pmatrix}~,
\quad P\in\iO\,\H\Bigr\}~,
\dbk[8pt]
&\D^4\equiv\weu4\H=
\Bigl\{\iO\,r\begin{pmatrix}\Id{\U} & 0 \\ 0& -\Id{\Ua}\end{pmatrix}~,
\quad r\in\RR\Bigr\}~,
\dbk[8pt]
&\D^\pmap=\D^0\oplus\D^2\oplus\D^4=
\Bigl\{\begin{pmatrix}K & 0 \\ 0 & K^\ddag\end{pmatrix}~,
\quad K\in\U\tn\Ul\Bigr\}~,
\dbk[8pt]
&\D^\mmap=\D^1\oplus\D^3=
\Bigl\{\begin{pmatrix}0 & P \\ P^\ddag & 0\end{pmatrix}~,
\quad P\in\U\tn\Uc \Bigr\}~.
\end{align*}
\end{proposition}
\proof
The Dirac map $\g:\H\to\End\W$ can be written as
$$\g:v\mapsto\begin{pmatrix}0&\sqrt2\,v \\ \sqrt2\,v^\ddag & 0\end{pmatrix}~,$$
whence the characterization of $\D^1$.
It immediately follows that
$\D^\pmap$ is constituted by diagonal-block elements,
while $\D^\mmap$ is constituted by off-diagonal-block elements.
The other characterizations can be checked by matrix calculations.\qed

\section{Clifford group and its subgroups}
\label{S:Clifford group and its subgroups}
\subsection{Clifford group}\label{s:Clifford group}

Let $\D^{\bul}:=\D\cap\Aut\W$ be the group of all invertible
elements in $\D$.
The Clifford group $\Cl\equiv\Cl(\W)$ is defined to be~\cite{Cr,Gr}
the subgroup of $\D^{\bul}$ under whose adjoint action $\H$ is stable.
In other terms, $\Phi\in\D^\bul$ is an element of $\Cl$ iff
$$\Ad[\Phi]v\equiv\Phi\,\g(v)\,\Phi^{-1}\in\g(\H)~,\quad\forall\:v\in\H~.$$
Using proposition~\ref{p:inverseofPhi} we write the adjoint action as
\begin{align*}
(\det\Phi)\,\Ad[\Phi]v &=
\begin{pmatrix}K && P \\[6pt] P^\ddag && K^\ddag \end{pmatrix}
\begin{pmatrix}0 && V \\[6pt] V^\ddag && 0\end{pmatrix}
\begin{pmatrix}X && Y \\[6pt] Y^\ddag && X^\ddag \end{pmatrix}=
\\[12pt]
&=\begin{pmatrix}
P\,V^\ddag\,X+K\,V\,Y^\ddag && P\,V^\ddag\,Y+K\,V\,X^\ddag
\\[8pt]
K^\ddag\,V^\ddag\,X+P^\ddag\,V\,Y^\ddag &&
K^\ddag\,V^\ddag\,Y+P^\ddag\,V\,X^\ddag
\end{pmatrix}~,
\end{align*}
where $V\equiv\sqrt2\,v$ and
\begin{align*}
& X\equiv (\det{\bar K})\,\td K{}^\lin-\bar P{}^\lin\,\tb K\,\td P{}^\lin~,
&& Y\equiv (\det P)\,\bar P{}^\lin-\td K{}^\lin\,P\,\bar K{}^\lin~,
\\[8pt]
& X^\ddag=(\det K)\,\bar K{}^\lin-\td P{}^\lin\,K\,\bar P{}^\lin~,
&& Y^\ddag=(\det\bar P)\,\td P{}^\lin-\bar K{}^\lin\,\tb P\,\td K{}^\lin~.
\end{align*}

\begin{lemma}\label{l:Cleitheroddoreven}
An element of $\D^\bul$ which belongs to the Clifford group
is necessarily either odd or even,
so that the Clifford group is the disjoint union $\Cl=\Cl^\pmap\cup\Cl^\mmap$
where $\Cl^\pmap\equiv\Cl\cap\D^\pmap$\,,
$\Cl ^\mmap\equiv\Cl\cap\D^\mmap$\,.
\end{lemma}
\proof
If $\Phi$ is in $\Cl$ then the $\U\tn\Ul$-component of $\Ad[\Phi]v$
vanishes for all $v\in\H$, namely
$$K\,V\,\tb Y=-P\,\tb V\,X~,\quad \forall\:V\in\H~.$$
Composing both sides with $\td V{}^\lin\,\td K{}^\lin$ on the left
and with $\td X{}^\lin$ on the right one finds
$$(\det{K})\,(\det{V})\,\tb Y\,\td X{}^\lin=
-(\det\Phi)(\det\bar K)\,\td V{}^\lin\,\td K{}^\lin\,P\,\tb V~.$$
Now the above equality is certainly fulfilled in the particular case
when $\det K=0$\,.
Suppose $\det K\neq0$ for the moment
(the other case will be considered later).
The left-hand side vanishes for all null elements $V\in\H$,
thus also $\td V{}^\lin\,\td K{}^\lin\,P\,\tb V$ vanishes
for all null vectors $V$\,;
it's not difficult to see that this implies $\td K{}^\lin\,P=0$\,,
which on turn implies $P=0$\,.
Summarizing, if $\Phi\in\Cl$ and $\det K\neq0$ then $P=0$\,.
By a similar argument,
composing the equation $K\,V\,\tb Y=-P\,\tb V\,X$
on the left by $\bar V{}^\lin\,\td P{}^\lin$
and on the right by $\bar Y{}^\lin$,
one finds that if $\Phi\in\Cl$ and $\det P\neq0$ then $K=0$\,.

The case which remains to be considered is that when $\det K=\det P=0$\,.
Since $\det P=\oh\,g(P,P)$\,, $P$ is an isotropic element of $\U\tn\Uc$,
and as such it is decomposable.
Similarly, $K$ is decomposable. Namely one can write
$$K=k\tn\l~,~~P=p\tn\bq~,~~V=s\tn\bs~,\qquad k,p,q,s\in\U,~\l\in\Ul~.$$
A little two-spinor algebra then yields
\begin{align*}
& P\,\tb V\,X+K\,V\,\tb Y=\be(\bar k,\bar p)\,\Bigl[\,
\bang{\l,q}\,|\bang{\l,s}|^2\,k\tn k^\fl
-\bang{\bl,\bq}\,|\e(s,q)|^2\,p\tn p^\fl\,\Bigr]~,
\\[6pt]
& \det\Phi=-\Tr(K\,\bar P{}^\lin\,\tb K\,\td P{}^\lin\,)=
|\e(k,p)|^2\,|\bang{\l,q}|^2~.
\end{align*}
Now one sees that in order that $\det\Phi\neq0$ one must have
$\bang{\l,q}\neq0$ and $\e(k,p)\neq0$\,.
Thus $k\tn k^\fl$ and $p\tn p^\fl$
are linearly independent elements of $\U\tn\Ul$
and, in order that $P\,\tb V\,X+K\,V\,\tb Y$ vanishes for all $V$,
one must have $\bang{\l,s}=\e(q,s)$ for all $s\in\U$\,,
which implies $\l=0$ and $q=0$ that is $K=0$ and $P=0$\,,
a contradiction.
Thus the case $\det K=\det P=0$ cannot yield an element $\Phi\in\Cl$\,.
\qed

\begin{proposition}~\\
$a)$~$\Cl^\pmap$ is the $7$-dimensional real submanifold of $\D^\pmap$
constituted of all elements in $\W\tn\Wl$ which are of the type
$$\sKM~,\quad K\in\U\tn\Ul\,,~\det K\in\RR\setminus\{0\}~.$$
$b)$~$\Cl^\mmap$ is the $7$-dimensional real submanifold of $\D^\mmap$
constituted of all elements in $\W\tn\Wl$ which are of the type
$$\sPM~,\quad P\in\U\tn\Uc\,,~\det P\in\RR\setminus\{0\}~.$$
\end{proposition}
\proof\\
$a)$~Let $\Phi=\sKM$\,, $K\in\U\tn\Ul$, $\det K\neq0$\,.
Then
$$(\det\Phi)\,\Ad[\Phi]v=
\begin{pmatrix}0 && (\det K)\,K\,V\,\bar K{}^\lin \\[8pt]
(\det\bar K)\tb K\,\tb V\,\,\td K{}^\lin && 0 \end{pmatrix}~,\quad
V\equiv\sqrt2\,v\in\H~.$$
For $\Ad[\Phi]v$ to be in $\H$,
the two non-zero entries of the above matrix must be in $\H\equiv\U\vh\Uc$
and in $\Ua\vh\Ul$, respectively.
Consider the $\U\tn\Uc$-entry.
Since $\bar V=V^\lin$ because $V$ is Hermitian, one finds
$$[(\det K)\,K\,V\,\bar K{}^\lin]^\alin=
(\det\bar K)\,K\,V\,\bar K{}^\lin~,$$
and $(\det K)\,K\,V\,\bar K{}^\lin$ is Hermitian
for all $V\in\H$ iff $\det K=\det\bar K$
(this argument gives the same result for the other non-zero entry).
\\[6pt]
$b)$~Let $\Phi=\sPM$\,, $P\in\U\tn\Uc$, $\det P=\oh\,g(P,P)\neq0$\,. Then
$$(\det\Phi)\,\Ad[\Phi]v=
\begin{pmatrix}0 && (\det P)\,P\,\tb V\,\bar P{}^\lin \\[8pt]
(\det\bar P)\tb P\,V\,\,\td P{}^\lin && 0 \end{pmatrix}~.$$
By the same argument as before, $\Phi\in\Cl$ iff $\det P=\det\bar P$\,.\qed

Now it is not difficult to show that any complex $2\times2$-matrix
with real determinant can be written as a product of Hermitian matrices.
Using this, one recovers a well-known result:
\begin{proposition}\label{p:Clmgenerated}
$\Cl$ is multiplicatively generated by $\H^\bul\subset\H$,
the subset of all elements in $\H$ with non-vanishing Lorentz pseudo-norm.
\end{proposition}
Namely any element of $\Cl$ can be written as
$$\Phi=v_1\,v_2\,\dots\,v_n~,\quad v_j\in\H,~g(v_j,v_j)\neq0~;$$
its inverse is
$$\Phi^{-1}=\frac1{\n(\Phi)}\,v_n\,\dots\,v_2\,v_1~,\quad
\n(\Phi):=g(v_1,v_1)\,g(v_2,v_2)\,\dots\,g(v_n,v_n)~.$$
Setting now $V_i\equiv\sqrt2\,v_i$ one has $\det V_i=\det\tb V_i=g(v_i,v_i)$\,,
hence
$$\n(\Phi)=\det\bigl(V_1\,\tb V_2\,V_3\,\tb V_4\,\dots\bigr)
=\mathop{\Pi}\limits_{i=1}^n \det(V_i)~.$$
Namely, if $\Phi=\sKM\in\Cl^\pmap$
then $\n(\Phi)=\det K=\det K^\ddag$\,;
if $\Phi=\sPM\in\Cl^\mmap$
then $\n(\Phi)=\det P=\det P^\ddag$\,.

\begin{remark}
Actually, it can be seen that any complex $2\times2$-matrix
with real determinant can be written as a product
of just \emph{three} Hermitian matrices
(but not, in general, of two matrices).
This implies that an element in $\Cl^\mmap$ can be written as $\sPM$
with $P=V_1\,V_2^\ddag\,V_3$\,,
and an element in $\Cl^\pmap$ can be written as $\sKM$
with $K=V_1\,V_2^\ddag\,V_3\,V_4^\ddag$\,,
$V_i\in\H^\bul$\,.
\end{remark}\bigbreak

The adjoint action of any $w\in\H$ on $\H$ is easily checked to be
the negative of the reflection 
through the hyperplane orthogonal to $w$\,.
It follows that $\Cl^\pmap$ is the subgroup of all elements in $\Cl$
whose adjoint action preserves the orientation of $\H$.
Moreover, the subgroup
$$\Cl^\up:=\{\Phi\in\Cl:\n(\Phi)>0\,\}$$
is constituted of all elements of $\Cl$
whose adjoint action preserves the time-orientation of $\H$.
Its representation as $\Phi=v_1\,v_2\,\dots\,v_n$
has an even number of spacelike factors
and any number of timelike factors.

\bigbreak
The unit element of $\Cl$ is $\id\in\D^\pmap\subset\D$.
Thus the Lie algebra of $\Cl$ is a 7-dimensional vector subspace
$$\Lie\Cl\subset\D^\pmap=\RR\oplus\weu2\H\oplus\weu4\H
\equiv \RR\,\id\oplus\hat\g(\weu2\H)\oplus\hat\g(\weu4\H)~.$$
Now observe that $\weu4\H$ is not contained in $\Lie\Cl$ since
$$t\in\RR\qRq\exp(t\,\eta^\#)=
\exp\begin{pmatrix} -\iO\,t\,\Id{\U} & 0 \\
0 & \iO\,t\,\Id{\Ua} \end{pmatrix}
=\begin{pmatrix}\eO^{-\iO\,t}\,\Id{\U} & 0 \\
0 & \eO^{\iO\,t}\,\Id{\Ua} \end{pmatrix}$$
is not in $\Cl$ because the two component endomorphsims
$\eO^{-\iO\,t}\,\Id{\U}\in\U\tn\Ul$ and $\eO^{\iO\,t}\,\Id{\Ua}\in\Ua\tn\Uc$
have non-real determinant.
Hence, just by a dimension argument, one finds
$$\Lie\Cl=\RR\oplus\weu2\H~.$$

\subsection{$\Pin$ and $\Spin$}\label{s:Pin and Spin}

If $\Phi\in\Cl$ and $a\in\RR\setminus\{0\}$
then $\Ad[a\,\Phi]=\Ad[\Phi]:\H\to\H$.
It is then natural to consider the subgroup
$$\Pin:=\{\Phi\in\Cl:\n(\Phi)=\pm1\}~,$$
which is multiplicatively generated by all elements in $\H$
whose Lorentz pseudo-norm is $\pm1$\,.
It has the subgroups
\begin{align*}
& \Spin:=\Pin^\pmap\equiv\Pin\cap\Cl^\pmap=\{\Phi\in\Cl^\pmap:\n(\Phi)=\pm1\}~,
\\[6pt]
& \Pin^\up:=\Pin\cap\Cl^\up=\{\Phi\in\Cl:\n(\Phi)=1\}~,
\\[6pt]
& \Spin^\up:=\Spin\cap\Cl^\up=\{\Phi\in\Cl^\pmap:\n(\Phi)=1\}~.
\end{align*}
These share the same Lie algebra
$$\weu2\H=\Lie\Pin=\Lie\Spin=\Lie\Pin^\up=\Lie\Spin^\up~.$$

\bigbreak
The automorphisms of $\U$ which have unit determinant constitute the group
$\SlG\equiv\SlG(\U)$\,;
thus
\begin{align*}
& \Cl^{\pmap\up}\equiv\Cl^\pmap\cap\Cl^\up
=\left\{\, \sKM\in\End\W : K\in\RR^+\times\SlG \,\right\}~,
\\[6pt]
& \Spin^\up=\left\{\, \sKM\in\End\W : K\in\SlG \,\right\}~.
\end{align*}
In particular, one has the isomorphism
$$\Spin^\up\leftrightarrow\SlG:\sKM\leftrightarrow K~.$$

Now remember that
\begin{align*}
& \hat\g(\weu2\H)=
\left\{\,\cliffordplusmatrix{A}\in\End\W : \Tr A=0\,\right\}~,
\\[8pt]
& \hat\g(\RR\oplus\weu2\H)=
\left\{\,\cliffordplusmatrix{A}\in\End\W : \Im\Tr A=0\,\right\}~;
\end{align*}
moreover $\End\U$ can be decomposed into the direct sum
of the subspace of all traceless endomorphism,
which is just $\Lie\SlG$\,,
and the subspace $\CC\,\id$ generated by the identity.
Then one has the Lie algebra isomorphisms
\begin{align*}
& \Lie\Cl=\Lie\Cl^{\pmap\up}=\RR\oplus\weu2\H ~\longrightarrow~
(\RR\,\id)\oplus\Lie\SlG~,
\\[6pt]
& \Lie\Pin=\Lie\Spin^\up=\weu2\H ~\longrightarrow~ \Lie\SlG~.
\end{align*}

\bigbreak
\begin{proposition}
Let
$$\Phi=\sKM\in\Spin~,~~v\in\H~,~~
\g(v)=\cliffordplusmatrix{V}\equiv
\left(\begin{smallmatrix}\sqrt2\,v&0\\
0&\sqrt2\,v^\ddag\end{smallmatrix}\right)~.$$
Then
$$\Ad[\Phi]\g(v)=\pm
\begin{pmatrix}0 && [K{\otimes}\bar K](V) \\[6pt]
\bigl([K{\otimes}\bar K](V)\bigr)^\ddag && 0\end{pmatrix}~,$$
where the $+$ sign holds iff $\Phi\in\Spin^\up$.
\end{proposition}
\proof
Remembering the previous results one finds
$$\Ad[\Phi]\g(v)=
\frac1{\det K}\begin{pmatrix}
0 && K\,V\,\bar K{}^\lin \\[6pt]
\bigl(K\,V\,\bar K{}^\lin\bigr)^\ddag && 0
\end{pmatrix}~.$$
Moreover
$$(K\,V\,\bar K{}^\lin)^\AAd
=K\Ii\sA\sB\,V^\BBd\,(\bar K{}^\lin)\iI\cB\cA
=K\Ii\sA\sB\,V^\BBd\,\bar K\Ii\cA\cB
=(K\tn\bar K)\Ii\AAd\BBd\,V^\BBd~.$$
\qed

Now remember (\Sec\ref{s:2-spinor groups and Lorentz group}) that the group
$\{K\tn\bar K:K\in\Aut(\U)\,\}$
is constituted of automorphisms of $\U\tn\Uc$
which preserve the splitting $\U\tn\Uc=\H\oplus\iO\,\H$
and the causal structure of $\H$.
Its subgroup
$\{K\tn\bar K:K\in\SlG(\U)\,\}$
coincides with $\Lor_+^\up(\H)$\,.
Thus one sees that the group isomorphism $\SlG\to\Spin^\up$
determines the \hbox{2-to-1} epimorphism $\Spin^\up\to\Lor_+^\up$\,.

\bigbreak
One also finds that $\Spin^\up$ is the subgroup of $\End\W$
preserving $(\g,\kO,g,\eta,\e)$ as well as time-orientation.
Let's review these properties in terms of two-spinors.

\bigbreak\noindent
$\bullet$~Obviously, $\Spin^\up$ preserves the splitting $\W=\U\oplus\Ua$.
If $\Phi=\sKM$\,, $K\in\SlG(\U)$\,, then $\td K=K^{-1}$\,,
so for $\psi\equiv(u,\chi),\psi'\equiv(u',\chi')\in\W$ one gets
\begin{align*}
\kO(\Phi\psi,\Phi\psi')&=
\kO\bigl((K\,u,\chi\,\bar K{}^{-1}),(K\,u',\chi'\,\bar K{}^{-1})=
\bang{\bch\,K^{-1},K\,u'}+\bang{\chi'\,\bar K{}^{-1},\bar K\,\bu}=\\[6pt]
&=\bang{\bch,u'}+\bang{\chi',\bu}=\kO(\psi,\psi')~.
\end{align*}

\smallbreak\noindent
$\bullet$~Since $K\tn\bar K:\U\tn\Uc\to\U\tn\Uc$
sends Hermitian tensors to Hermitian tensors
and anti-Hermitian tensors to anti-Hermitian tensors,
it preserves the splitting $\U\tn\Uc=\H\oplus\iO\,\H$.
Also, remember that $K\tn\bar K=\Ad[\Phi]$\,.

\smallbreak\noindent
$\bullet$~$K\tn\bar K=\Ad[\Phi]\in\Lor_+^\up(\H)$\,,
the subgroup of the Lorentz group which preserves orientation
and time-orientation.

\smallbreak\noindent
$\bullet$~$\Phi$ preserves the Dirac map $\g$\,. In fact if $y\in\H$ then
\begin{align*}
&\g[y]=\begin{pmatrix}0 && \sqrt2\,y \\ \sqrt2\,y^\ddag && 0\end{pmatrix}~,
\quad y^\ddag\equiv\tb y=\td y{}^\lin~,
\\[8pt]
& \Ad[\Phi]\g[y]=
\begin{pmatrix}0 && \sqrt2\,[K\tn\bar K]\,y \\
\sqrt2\,([K\tn\bar K]\,y)^\ddag && 0\end{pmatrix}
=\g\bigl[[K\tn\bar K]\,y\bigr]~.
\end{align*}

\smallbreak\noindent
$\bullet$~If $K\in\SlG$ then $K$ preserves any simplectic form $\e\in\weu2\Ul$.
Hence $\Phi\equiv\sKM\in\Spin^\up$ preserves the corresponding
simplectic form $(\e,\be^\#)\in\weu2\Wl$ and charge conjugation.

\section{Spinors and particle momenta}
\label{S:Spinors and particle momenta}
\subsection{Particle momentum in two-spinor terms}
  \label{s:Particle momentum in two-spinor terms}

It has already been observed (\Sec\ref{s:2-spinors and Minkowski space})
that any future-pointing non-spacelike element in $\H$
can be written in the form
$$u\tn\bu+v\tn\bv~,\quad u,v\in\U~.$$
If $u$ and $v$ are not proportional to each other, that is $\e(u,v)\neq0$\,,
then the above expression is a timelike future-pointing vector;
if $\e(u,v)\neq0$\,, then it is a null vector.
Future-pointing elements in $\H$ are a contravariant,
``conformally invariant'' version of \emph{classical particle momenta}
(translation to a scaled and/or covariant version, when needed,
will be effortless).

Let $\K$ and $\N$ be the subsets of $\H$ constituted
of all future-pointing timelike vectors
and of all future-pointing null vectors, respectively;
moreover, set $\J:=\K\cup\N$
(all these sets do not contain the zero element).
Consider now the real quadratic maps
\begin{align*}
&\tdpO:\U\times\U\to\J:(u,v)\mapsto\osq\,(u\tn\bu+v\tn\bv)~,
\\[6pt]
&\pO:\W\cong\U\times\Ua\to\J:
(u,\chi)\mapsto\osq\,(u\tn\bu+\bch^\#\tn\chi^\#)~.
\end{align*}
When a normalized symplectic form $\e\in\weu2\Ul$ is \emph{fixed},
$\tdpO$ and $\pO$ are essentially the same objects,
as one can represent a given element
$\osq\,(u\tn\bu+v\tn\bv)$ of $\J$
by writing $v\tn\bv$ as $(\bch\tn\chi)^\#$\,;
here, $u,v\in\U$, $\chi\in\Ua$.
In such case I'll set
\begin{align*}
& v:=-\bch^\# \quad\iff\quad \chi=\bv^\fl~,\\[6pt]
\Rightarrow\quad
&\bang{\bch,u}=\bang{v^\fl,u}=\e(v,u)~,\quad
\bang{\chi,\bu}=\bang{\bv^\fl,\bu}=\be(\bv,\bu)~.
\end{align*}

If $p=\pO(u,\chi)\equiv\tdpO(u,v)$ then we'll use the shorthands
\begin{align*}
& \m^2:=g(p,p)=2\,|\e(u,v)|^2=2\,|\bang{\bch,u}|^2~,\\[6pt]
& h:=\frac{\sqrt2}\m\,\bar p^\fl
=\frac1{|\bang{\bch,u}|}\,(\bu^\fl\tn u^\fl+\chi\tn\bch)~.
\end{align*}
Then, $h$ can be seen as an $\e$-normalized Hermitian metric on $\U$.

\begin{proposition}\label{p:pgeneratingcouple}
Let $(u,\chi)\equiv(u,\bv^\fl)\in\W$\,, $\bang{\bch,u}\neq0$\,;
let $p\in\K$\,.
Then, the following conditions are equivalent:
\begin{enumerate}
\item[{\bf i)}]
$p=u\tn\bu+(\bch\tn\chi)^\#$\,,
\item[{\bf ii)}]
$\g[p](u,\chi)=\m\,\bigl(\eO^{-\iO\th}u,\eO^{\iO\th}\,\chi\bigr)$\,,~
$\th\in\RR$\,,
\item[{\bf iii)}]
$\bh^\fl(u)=\eO^{\iO\th}\,\chi$\,,
\item[{\bf iv)}]
$h^\#(\chi)=\eO^{-\iO\th}\,u$\,,
\item[{\bf v)}]
$h(\bu,v)=0$ and $|\bang{\bch,u}|=h(\bu,u)$\,,
\item[{\bf v')}]
$h(\bu,v)=0$ and $|\bang{\bch,u}|=h(\bv,v)$\,,
\end{enumerate}
where $\m$ and $h$ are defined in terms of $(u,\chi)$ as above.
\end{proposition}
\proof
By straightorwaed calculations one sees that condition {\bf i}
implies conditions {\bf ii}, {\bf iii}, {\bf iv}, {\bf v} and {\bf v'}.
Moreover:
\smallbreak\noindent
{\bf (\:ii}~$\Leftrightarrow$~{\bf iii\,)\,:}~%
It follows from
$\g[\t](u,\chi)=\osq\,\g[\bh^\#](u,\chi)
=\bigl(h^\#(\chi),\bh^\fl(u)\bigr)$\,.

\smallbreak\noindent
{\bf (\:iii}~$\Leftrightarrow$~{\bf iv\,)\,:}~%
If $\bh^\fl(u)=\eO^{\iO\th}\,\chi$ then
$u=h^\#(\bh^\fl(u))=h^\#(\eO^{\iO\th}\,\chi)=\eO^{\iO\th}\,h^\#(\chi)$\,.\\
Similarly, if $h^\#(\chi)=\eO^{-\iO\th}\,u$ then
$\chi=\bh^\fl(h^\#(\chi))=\bh^\fl(\eO^{-\iO\th}\,u)
=\eO^{-\iO\th}\,\bh^\fl(u)$\,.

\smallbreak\noindent
{\bf (\:iv}~$\Rightarrow$~{\bf v\,)\,:}~%
$h(\bu,v)=\bang{h^\fl(\bu),-\bch^\#}=-\bang{\eO^{-\iO\th}\,\bch,\bch^\#}
=\eO^{-\iO\th}\,\e^\#(\bch,\bch)=0$\,.\\
Moreover
$h(\bu,u)=\bang{\bh^\fl(u),\bu}=\bang{\eO^{\iO\th}\,\chi,\bu}
=\bang{\bch,u}\,\bang{\chi,\bu}/|\bang{\bch,u}|=|\bang{\bch,u}|$\,.

\smallbreak\noindent
{\bf (\:v}~$\Rightarrow$~{\bf iv\,)\,:}~%
From $0=h(\bu,v)=\bang{h^\fl(\bu),-\bch^\#}=-\e^\#(\bch,h^\fl(\bu))$
one has $\bch=c\,h^\fl(\bu)$\,, $c\in\CC$\,.
Then $\bang{\bch,u}=c\,h(\bu,u)=c\,|\bang{\bch,u}|\qRq c=\eO^{\iO\th}$\,.

\smallbreak\noindent
{\bf (\:v}~$\Rightarrow$~{\bf v'\,)\,:}~%
From {\bf iv} (equivalent to {\bf v}) one has
$h(\bv,v)=\bang{h,\chi^\#\tn\bch^\#}=\bang{h^\#,\chi\tn\bch}
=\bang{h^\#(\chi),\bch}=\eO^{-\iO\th}\,\bang{\bch,u}=|\bang{\bch,u}|$\,,
hence also $h(\bv,v)=|\bang{\bch,u}|$\,.

\smallbreak\noindent
{\bf (\:v'}~$\Rightarrow$~{\bf iv\,)\,:}~%
As in {\bf v}~$\Rightarrow$~{\bf iv} one has $\bch=c\,h^\fl(u)$\,, $c\in\CC$\,,
or $u=\frac1{\bar c}\,h^\#(\chi)$\,.
Then, from
$\bang{\bch,u}=\bang{\bch,\frac1{\bar c}\,h^\#(\chi)}
=\frac1{\bar c}\,h^\#(\chi,\bch)=\frac1{\bar c}\,h(\bv,v)$
one has $\bar c=\eO^{-\iO\th}$ \ie\ $c=\eO^{\iO\th}$\,.

\smallbreak\noindent
{\bf (\:v}~$\Rightarrow$~{\bf i\,)\,:}~%
Using also {\bf v'} (already seen to be equivalent to {\bf v})
one sees that the couple $(\z_u\,,\z_v)\equiv(u,v)/\sqrt{|\bang{\bch,u}|}$
is an $h$-orthonormal basis of $\U$\,;
hence
$h^\#=\bze_u\tn\z_u+\bze_v\tn\z_v
=\frac1{|\bang{\bch,u}|}\,\bigl(\bu\tn u+\bv\tn v\bigr)~.$
Condition {\bf i} then follows.
\qed

\subsection{Bundle structure of 4-spinor space over momentum space}
  \label{s:Bundle structure of 4-spinor space over momentum space}

The previous results show that the restriction
$\pO:\W\setminus\{0\}~\longrightarrow~\J$
is surjective.
Since the Lorentz ``length'' of $\pO(u,\chi)$ is $\sqrt2\,|\bang{\bch,u}|$
one sees that the subset of all elements in $\W$
which project onto $\N$ is the 6-dimensional real submanifold
$$\W^0:=\pO^{-1}(\N)=
\bigl\{(u,\chi)\in\W{\setminus}\{0\}:\bang{\bch,u}=0\bigr\}
\subset\W~.$$
The subset of all elements in $\W$ which project onto $\K$ is
the open submanifold
$$\W^\bp:=\pO^{-1}(\K)=
\bigl\{(u,\chi)\in\W:\bang{\bch,u}\neq0\bigr\}~,$$
and one has
$$\W{\setminus}\{0\}=\W^0\cup\W^\bp~.$$
Moreover, consider the subsets $\W^+,\W^-\subset\W^\bp$ defined to be
$$\W^\pm:=\bigl\{(u,\chi)\in\W:\bang{\bch,u}\in\RR^\pm\bigr\}~.$$
Recalling condition {\bf ii} of proposition~\ref{p:pgeneratingcouple} one has
$$\g[\pO\psi]\psi=\m\,(\eO^{-\iO\th}u,\eO^{\iO\th}\,\chi)~,$$
which holds for every $\psi\equiv(u,\chi)\in\W$
(if $\psi\in\W^0$ then $\m=0$).
In particular
$$\W^\pm=\bigl\{\: \psi\equiv(u,\chi)\in\W{\setminus}\{0\} :
\g[\pO\psi]\psi=\pm\m\,\psi~,\m\equiv|\bang{\bch,u}|\:\bigr\}~.$$

Next, consider the subset
$$\td\W{}^\bp:=\{(u,v):\e(u,v)\neq0\}\subset\U\times\U~,$$
and note that when a normalized symplectic form $\e\in\weu2\Ul$ is fixed,
$\td\W{}^\bp$ can be identified with $\W^\bp$ via the correspondence
$\bv^\fl\leftrightarrow\chi$\,.
$\td\W{}^\bp$ is a fibred set over $\K$\,;
for each $p\in\K$, the fibre of $\td\W{}^\bp$ over $p$ is the subset
$$\td\W_{\!\!p}^\bp:=\tdpO{}^{\sst-1}(p)=\bigl\{
(u,v)\in\td\W{}^\bp:\osq\,(u\tn\bu+v\tn\bv)=p \bigr\}~.$$

\begin{proposition}\label{p:uvU2}
$\tdpO:\td\W{}^\bp\to\K$
is a trivializable principal bundle with structure group $\Ug(2)$\,.
\end{proposition}
\proof
Let $p=\tdpO(u,v)=\tdpO(u',v')$\,.
From proposition~\ref{p:pgeneratingcouple} one then sees that
$(u,v)$ and $(u',v')$ are orthonormal bases of $\U$
relatively to the Hermitian metric
$h\equiv\sqrt2\,\bar p^\fl/\m$.
Then there exists a unique transformation $K\in\Ug(\U,h)$ such that
$$u'=K(u)~,\quad v'=K(v)~;$$
hence,
$\td\W_{\!\!p}^\bp$ is a group-affine space, with derived group $\Ug(2)$\,.

Let now $(\zeA)$ be an $\e$-normalized basis of $\U$ and $(\t_\l)$
the associated Pauli frame.
For each $p\in\K$ let $L_p\in\Lor_+^\up(\H)$ be the boost
such that $L_p\t_0=p/\m$\,, where $\m^2\equiv g(p,p)$\,;
up to sign there is a unique $B_p\in\SlG(\U)$ such that
$L_p=B_p\tn\bar B_p$\,,
and a consistent smooth way of choosing one such $B_p$ for each $p$
can be fixed.
It turns out that the basis $\bigl(\sqrt\m\,B_p\zeA\bigr)$
is orthonormal relatively to $\sqrt2\,\bar p^\fl/\m$
seen as a Hermitian metric on $\U$,
hence $\tdpO(\sqrt\m\,B_p\z_1\,,\sqrt\m\,B_p\z_2)=p$\,.
In this way one selects an ``origin'' element in each fibre of $\tdpO$\,,
so getting a trivialization $\td\W{}^\bp\to\K\times\Ug(2)$\,.
\qed

Using a little two-spinor algebra it is not difficult to prove:
\begin{proposition}\label{p:Kuchiexpr}
Let $\psi,\psi'\in\W^\bp$,
$\psi\equiv(u,\chi)$\,, $\psi'\equiv(u',\chi')$\,;
let $K\in\Aut\U$ be the unique automorphism of $\U$ such that
$$K\,u=u~,\quad K\,\bch^\#=\bch'{}^\#~.$$
Then
$$K=\frac1{\bang{\bch,u}^2}\,\bigl[
\bang{\bch,u'}\,u\tn\bch-\e^\#(\bch,\bch')\,u\tn u^\fl
+\e(u,u')\,\bch^\#\tn\bch+\bang{\bch',u}\,\bch^\#\tn u^\fl \bigr]~.$$
Moreover, one has
$$\chi'=K^\ddag\,\chi~.$$
Conversely,
the conditions $u'=Ku$ and $\chi'=K^\ddag\chi$ determine $K$ uniquely.
\end{proposition}

\bigbreak
The above expression for $K$ is invariant
relatively to the transformation $\e\mapsto\eO^{\iO\,\th}\e$\,;
hence, $K$ is independent of the particular normalized
symplectic form $\e$ chosen.

When a normalized $\e\in\weu2\Ul$ is given,
one has the real vector bundle isomorphism
$\W^\bp\leftrightarrow\td\W{}^\bp:(u,v)\leftrightarrow(u,\bv^\fl)$.
Through this correspondence,
$\W^\bp\to\K$ turns out to be a trivializable principal bundle
with structure group $\Ug(2)$\,.
If $\psi,\psi'\in\W_{\!\!p}^\bp$\,, let
$$(K)=c \begin{pmatrix}\hm a & \bar b~ \\ -b & \bar a\end{pmatrix}
\in\Ug(2)~,\quad
a,b,c\in\CC:|a|^2+|b|^2=|c|^2=1~,$$
be the matrix
of $K\in\Aut\U$ sending $\psi$ to $\psi'$
(according to proposition~\ref{p:Kuchiexpr})
relatively to the basis $(u,v)$\,.
Then
$$\begin{cases}
u'=c\,(a\,u-b\,v)~,\\[6pt]
v'=c\,(\bar b\,u+\bar a\,v)~,
\end{cases}
\qquad\quad\Longleftrightarrow\quad\qquad
\begin{cases}
u'=c\,(a\,u+b\,\bch^\#)~,\\[6pt]
\chi'=\bar c\,(a\,\chi+b\,\bu^\fl)~.
\end{cases}$$
If you take a different normalized symplectic form $\e\to\eO^{\iO\,\th}\e$\,,
then $K$ does not change,
while the corresponding matrix $(K)\in\Ug(2)$ changes
according to 
\hbox{$c\to c$}\,, \hbox{$a\to a$}\,, \hbox{$b\to\eO^{\iO\,\th}b$}\,.

\bigbreak
The above $\Ug(2)$-action does not preserve $\W^\pm\subset\W^\bp$.
In fact it's straightforward to prove:
\begin{proposition}
Let $\psi,\psi'\in\W_{\!\!p}^+$ (resp.\ $\psi,\psi'\in\W_{\!\!p}^-$),
$\psi\equiv(u,\chi)$\,, $\psi'\equiv(u',\chi')$\,;
let $K$ be the unique automorphism of $\U$ such that
$Ku=u$\,, $K^\ddag\chi=\chi'$\,.
Then $K\in\SU(\U,h)$\,, where $h\equiv\sqrt2\,\bar p^\fl/\m$\,.
\end{proposition}
Hence, $\W^+\to\K$ and $\W^-\to\K$
turn out to be trivializable principal bundles,
with structure group $\SU(2)$\,.
\vfill\newpage


\end{document}